\documentclass[review, sort&compress]{elsarticle}

\usepackage{lineno,hyperref}
\modulolinenumbers[5]

\usepackage{subcaption}
\usepackage{amsmath}
\usepackage{algorithm2e}
\usepackage{listings}
\usepackage{color}
\usepackage{tcolorbox}
\usepackage{geometry}
\journal{ArXiv}

\begin{document}

\begin{frontmatter}

  \title{Hydrograph peak-shaving using a graph-theoretic algorithm for placement of
    hydraulic control structures}
% \tnotetext[mytitlenote]{Fully documented templates are available in the elsarticle package on \href{http://www.ctan.org/tex-archive/macros/latex/contrib/elsarticle}{CTAN}.}

\author[civil]{Matthew~Bartos\corref{cor1}}
\ead{mdbartos@umich.edu}

\author[civil]{Branko~Kerkez}
\ead{bkerkez@umich.edu}

\cortext[cor1]{Corresponding author}
\address[civil]{Department of Civil and Environmental
  Engineering, University of Michigan, Ann Arbor, Michigan, USA.}

\begin{abstract}
  The need to attenuate hydrograph peaks is central to the design of stormwater
  and flood control systems. However, few guidelines exist for siting hydraulic
  control structures such that system-scale benefits are maximized. This study
  presents a graph-theoretic algorithm for stabilizing the hydrologic response
  of watersheds by placing controllers at strategic locations in the drainage
  network. This algorithm identifies subcatchments that dominate the peak of the
  hydrograph, and then finds the ``cuts'' in the drainage network that maximally
  isolate these subcatchments, thereby flattening the hydrologic response.
  Evaluating the performance of the algorithm through an ensemble of
  hydrodynamic simulations, we find that our controller placement algorithm
  produces consistently flatter discharges than randomized controller
  configurations---both in terms of the peak discharge and the overall variance
  of the hydrograph. By attenuating flashy flows, our algorithm provides a
  powerful methodology for mitigating flash floods, reducing erosion, and
  protecting aquatic ecosystems. More broadly, we show that controller placement
  exerts an important influence on the hydrologic response and demonstrate that
  analysis of drainage network structure can inform more effective stormwater
  control policies.
\end{abstract}

\begin{keyword}
Control systems, controller placement, graph theory, stormwater control
\end{keyword}

\end{frontmatter}

\clearpage

\section*{Highlights}

\begin{itemize}
    \item New algorithm for placing hydraulic control structures in drainage
      networks.

    \item Attenuates downstream hydrograph by de-synchronizing tributary flows.

    \item Algorithm is fast and requires only digital elevation data.
\end{itemize}

% \linenumbers

\section{Introduction}
\label{sec:intro}

In the wake of rapid urbanization, aging infrastructure and a changing climate,
effective stormwater management poses a major challenge for cities worldwide
\cite{Kerkez_2016}. Flash floods are one of the largest causes of natural
disaster deaths in the developed world \cite{Doocy_2013}, and often occur when
stormwater systems fail to convey runoff from urban areas \cite{Wright_2017}. At
the same time, many cities suffer from impaired water quality due to inadequate
stormwater control \cite{walsh_2005}. Flashy flows erode streambeds, release
sediment-bound pollutants, and damage aquatic habitats \cite{walsh_2005,
  booth_1997, finkenbine_2000, wang_2001}, while untreated runoff may trigger
fish kills and toxic algal blooms \cite{sahagun_2013, wines_2014}. Engineers
have historically responded to these problems by expanding and upsizing
stormwater control infrastructure \cite{rosenberg_2010}. However, larger
infrastructure frequently brings adverse side-effects, such as dam-induced
disruption of riparian ecosystems \cite{dams_and_development_2000}, and erosive
discharges due to overdesigned conveyance infrastructure \cite{Kerkez_2016}. As
a result, recent work has called for the replacement of traditional peak
attenuation infrastructure with targeted solutions that better reduce
environmental impacts \cite{arora_2015, Hawley_2016}.

As the drawbacks of oversized stormwater infrastructure become more apparent,
many cities are turning towards decentralized stormwater solutions to regulate
and treat urban runoff while reducing adverse impacts. Green infrastructure, for
instance, uses low-impact rain gardens, bioswales, and green roofs to condition
flashy flows and remove contaminants \cite{coffman_1999, strecker_2000,
  askarizadeh_2015}. \textit{Smart} stormwater systems take this idea further by
retrofitting static infrastructure with dynamically controlled valves, gates and
pumps \cite{Kerkez_2016, Bartos_2018, Mullapudi_2017, Mullapudi_2018}. By
actuating small, distributed storage basins and conveyance structures in
real-time, \textit{smart} stormwater systems can halt combined sewer overflows
\cite{Montestruque_2015}, mitigate flooding \cite{Kerkez_2016}, and improve
water quality at a fraction of the cost of new construction \cite{Kerkez_2016,
  Bartos_2018}. While decentralized stormwater management tools show promise
towards mitigating urban water problems, it is currently unclear how these
systems can be designed to achieve maximal benefits at the watershed scale.
Indeed, some research suggests when stormwater control facilities are not
designed in a global context, local best management practices can lead to
adverse system-scale outcomes---in some cases inducing downstream flows that are
more intense than those produced under unregulated conditions
\cite{Emerson_2005, petrucci_2013}.

Thus, as cities begin to experiment with decentralized stormwater control, the
question of \textit{where} to place control structures becomes crucial. While
many studies have investigated the ways in which active control can realize
system-scale benefits (using techniques like feedback control \cite{wong_2018},
market-based control \cite{Montestruque_2015}, or model-predictive control,
\cite{gelormino_1994, mollerup_2016}), the location of control structures within
the drainage network may serve an equally important function. Hydrologists have
long recognized the role that drainage network topology plays in shaping
hydrologic response \cite{kirkby_1976, gupta_1986, gupta_1988, mesa_1986,
  marani_1991, troutman_1985, Mantilla_2011, Tejedor_2015a, Tejedor_2015b}. It
follows that strategic placement of hydraulic control structures can shape the
hydrograph to fulfill operational objectives, such as maximally flattening flood
waves and regulating erosion downstream. To date, however, little research has
been done to assess the problem of optimal placement of hydraulic control
structures in drainage networks:

\begin{itemize}
\item Recent studies have investigated optimal placement of green infrastructure
  upgrades like green roofs, rain tanks and bioswales \cite{Zellner_2016,
    schubert_2017, yao_2015, zhang_2015, norton_2015, meerow_2017, schilling_2008}. However,
  these studies generally focus on quantifying the potential benefits of green
  infrastructure projects through representative case studies
  \cite{Zellner_2016, schubert_2017, yao_2015, zhang_2015}, and do not intend to
  present a generalized framework for placement of stormwater control
  structures. As a result, many of these studies focus on optimizing multiple
  objectives (such as urban heat island mitigation \cite{norton_2015}, air
  quality \cite{meerow_2017}, or quality of life considerations
  \cite{schilling_2008}), or use complex socio-physical models and optimization
  frameworks \cite{Zellner_2016}, making it difficult to draw general
  conclusions about controller placement in drainage networks.

\item Studies of pressurized water distribution networks have investigated the
  related problems of valve placement \cite{cattafi_2011, creaco_2010}, sensor
  placement \cite{Perelman_2013}, subnetwork vulnerability assessment
  \cite{Yazdani_2011}, and network sectorization \cite{Tzatchkov_2008,
    Hajebi_2015}. While these studies provide valuable insights into the ways
  that complex network theory can inform drinking water infrastructure design,
  water distribution networks are pressure-driven and cyclic, and are thus
  governed by different dynamics than natural drainage networks, which are
  mainly gravity-driven and dendritic.

\item Inspiration for the controller placement problem can be drawn from recent
  theoretical work into the controllability of complex networks. These studies
  show that the control properties of complex systems ranging from power grids
  to gene expression pathways are inextricably linked with topological
  properties of an underlying network representation \cite{liu_2016}. The
  location of driver nodes needed for complete controllability of a linear
  system, for instance, can be determined from the maximum matching of a graph
  associated with that system's state space representation \cite{Liu_2011}. For
  systems in which complete control of the network is infeasible, the relative
  performance of driver node configurations can be measured by detecting
  controllable substructures \cite{Ruths_2014}, or by leveraging the concept of
  ``control energy'' from classical control theory \cite{Summers_2014, yan_2012,
    yan_2015, shirin_2017}. While these studies bring a theoretical foundation
  to the problem of controller placement, they generally assume linear system
  dynamics, and may thus not be well-suited for drainage networks, which are
  driven by nonlinear runoff formation and channel routing processes.
\end{itemize}

Despite the critical need for system-scale stormwater control, there is to our
knowledge no robust theoretical framework for determining optimal placement of
hydraulic control structures within drainage networks. To address this knowledge
gap, we formulate a new graph-theoretic algorithm that uses the network structure
of watersheds to determine the controller locations that will maximally
``de-synchronize'' tributary flows. By flattening the discharge hydrograph, our
algorithm provides a powerful method to mitigate flash floods and curtail water
quality impairments in urban watersheds. Our approach is distinguished by the
fact that it is theoretically-motivated, and links the control of stormwater
systems with the underlying structure of the drainage network. The result is a
fast, generalized algorithm that requires only digital elevation data for the
watershed of interest. More broadly, through our graph-theoretic framework we
show that network structure plays a dominant role in the control of drainage
basins, and demonstrate how the study of watersheds as complex networks can
inform more effective stormwater infrastructure design.

\section{Algorithm description}
\label{sec:meth}

Flashy flows occur when large volumes of runoff arrive synchronously at a given
location in the drainage network. If hydraulic control structures are placed at
strategic locations, flood waves can be mitigated by ``de-synchronizing''
tributary flows before they arrive at a common junction. With this in mind, we
introduce a controller placement algorithm that minimizes flashy flows by
removing regions of the drainage network that contribute disproportionately to
synchronous flows at the outlet. In our approach, the watershed is first
transformed into a directed graph consisting of unit subcatchments (vertices)
connected by flow paths (edges). Next, critical regions are identified by
computing the catchment's \textit{width function} (an approximation of the
distribution of travel times to the outlet), and then weighting each vertex in
the network in proportion to the number of vertices that share the same travel
time to the outlet. The weights are used to compute a \textit{weighted
  accumulation} score for each vertex, which sums the weights of every possible
subcatchment in the watershed. The graph is then partitioned recursively based
on this weighted accumulation score, with the most downstream vertex of each
partition representing a controller location.

\begin{figure*}[htb!] \centering
  \includegraphics[width=\textwidth]{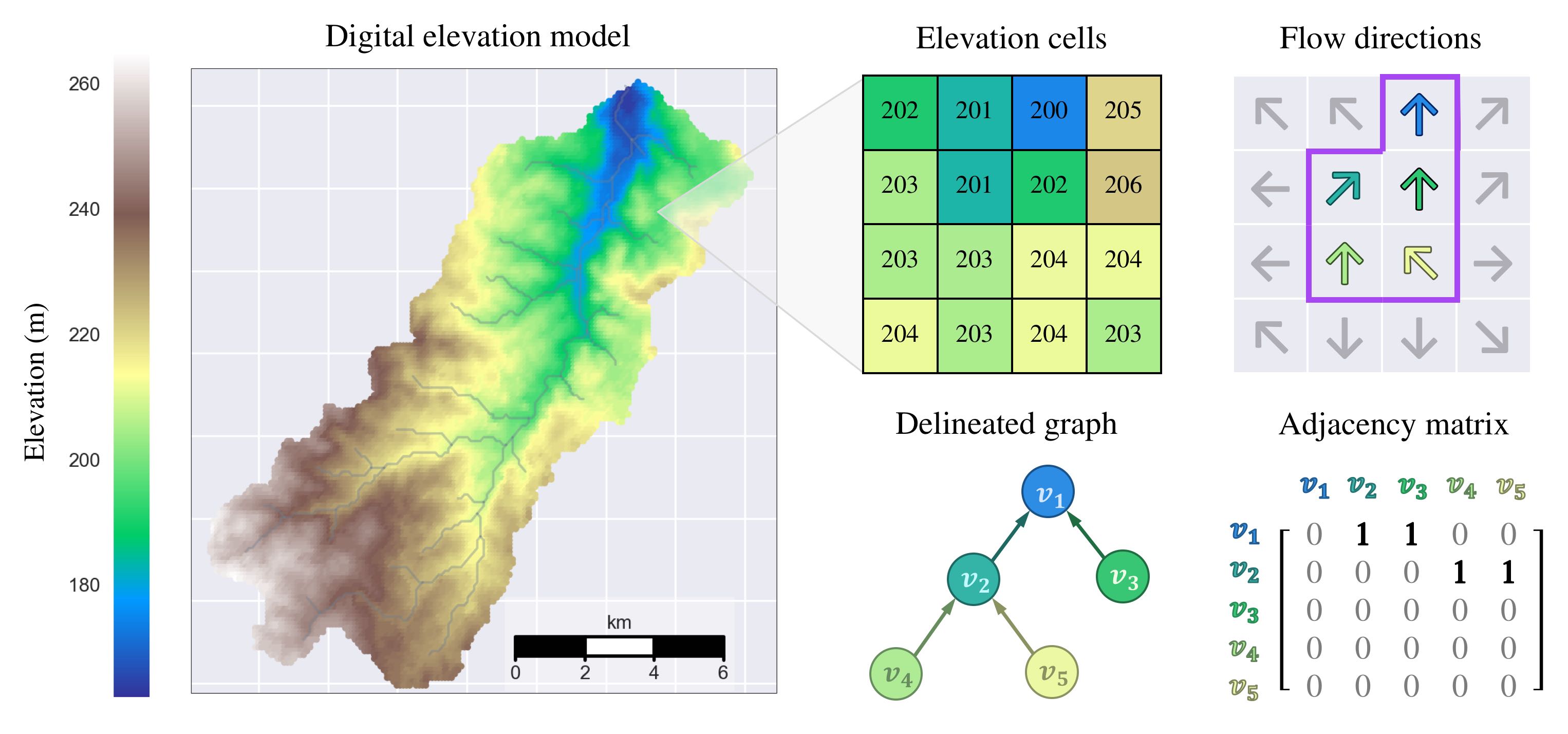}
  \caption[Watershed elevation and accumulation]{Left panel: Digital elevation
    model (DEM) of a watershed with river network highlighted. Right panel (from
    left to right, top to bottom): (i) DEM detail (colors not to scale); (ii)
    flow directions; (iii) delineated subcatchment graph; (iv) adjacency matrix
    representation of graph.}
  \label{fig:elev_and_river}
\end{figure*}

\subsection{Definitions}

\textbf{Graph representation of a watershed:} Watersheds can be represented as
directed graphs, in which subcatchments (vertices or cells) are connected by
elevation-dependent flow paths (edges). The directed graph can be formulated
mathematically as an adjacency matrix, $A$, where for each element $a_{i,j}$,
$a_{i,j} \neq 0$ if there exists a directed edge connecting vertex $v_i$ to
$v_j$, and conversely, $a_{i,j} = 0$ if there does not exist a directed edge
connecting vertex $v_i$ to $v_j$. Nonzero edge weights can be specified to
represent travel times, distances, or probabilities of transition between
connected vertices. Flow paths between adjacent cells are established using a
routing scheme, typically based on directions of steepest descent (see Figure
\ref{fig:elev_and_river}).

In this study, we determine the connectivity of the drainage network using a
\textit{D8 routing} scheme \cite{o_callaghan_1984}. In this scheme, elevation
cells are treated as vertices in a 2-dimensional lattice (meaning that each
vertex $v_i$ is surrounded by eight neighbors $\mathcal{N}_i$). A directed link
is established from vertex $v_i$ to a neighboring vertex $v_j$ if the slope
between $v_i$ and $v_j$ is steeper than the slope between $v_i$ and all of its
other neighbors $\mathcal{N}_i \setminus v_j$ (where $v_j$ has a lower elevation
than $v_i$). The \textit{D8 routing} scheme produces a directed acyclic graph
where the indegree of each vertex is between 0 and 8, and the outdegree of each
vertex is 1, except for the watershed outlet, which is zero. It should be noted
that other schemes exist for determining drainage network structure, such as the
\textit{D-infinity} routing algorithm, which better resolves drainage directions
on hillslopes \cite{tarboton_1997}. However, because the routing scheme is not
essential to the construction of the algorithm, we focus on the simpler
\textit{D8} routing scheme for this study. Similarly, to simplify the
construction of the algorithm, we will assume that the vertices of the watershed
are defined on a regular grid, such that the area of each unit subcatchment is
equal. Figure \ref{fig:elev_and_river} shows the result of delineating a river
network from a digital elevation model (left), along with an illustration of the
underlying graph structure and adjacency matrix representation (right).

\textbf{Controller}: In the context of this study, a controller represents any
structure or practice that can regulate flows from an upstream channel segment
to a downstream one. Examples include retention basins, dams, weirs, gates and
other hydraulic control structures. These structures may be either passively or
actively controlled. For the validation assessment presented later in this
paper, we will examine the controller placement problem in the context of
\textit{volume capture}, meaning that controllers are passive, and that they are
large enough to completely remove flows from their upstream contributing areas.
However, the algorithm itself does not require the controller to meet these
particular conditions.

Mathematically, we can think of a controller as a cut in the graph that
removes one of the edges. This cut halts or inhibits flows across the affected
edge. Because the watershed has a dendritic structure, any cut in the network
will split the network into two sub-trees: (i) the delineated region upstream of
the cut, and (ii) all the vertices that are not part of the delineated region.
Placing controllers is thus equivalent to removing branches (subcatchments) from
a tree (the parent watershed).

\textbf{Delineation}: Delineation returns the set of vertices upstream of a
target vertex. In other words, this operation returns the contributing area of
vertex $v_i$. Expressed in terms of the adjacency matrix:

\begin{equation}
    \begin{split} V_{d}(A, v_i) = \{ v_j \in V | (A^n)_{ij} \not = 0 \text{ for
some } n \leq D \}
    \end{split}
\end{equation}

Where $A^n$ is the adjacency matrix $A$ raised to the $n^{th}$ power, $i$ is the
row index, $j$ is the column index, $V$ is the vertex set of $A$, and $D$ is
the graph diameter. Note that $(A^n)_{ij}$ is nonzero only if vertex $v_j$ is
located within an n-hop neighborhood of vertex $v_i$.

\textbf{Pruning}: Pruning is the complement of delineation. This operation
returns the vertex set consisting of all vertices that are not upstream of the
current vertex.

\begin{equation} V_p(A, v_i) = V \setminus V_{d}(A, v_i)
\end{equation}

Subgraphs induced by the delineated and pruned vertex sets are defined as follows:

\begin{equation}
    \begin{split} A_d(A, v_i) = A(G[V_d]) \\ A_p(A, v_i) = A(G[V_p])
    \end{split}
\end{equation}

Where $A(G[V])$ represents the adjacency matrix of the subgraph induced by the
vertex set $V$.

\clearpage

\textbf{Width function}: The width function describes the distribution of travel
times from each upstream vertex to some downstream vertex, $v_i$\footnote{The
  width function $H(x)$ was originally defined by Shreve (1969) to yield the
  number of links in the network at a topological distance $x$ from the outlet
  \cite{shreve_1969}. Because travel times may vary between hillslope and
  channel links, we present a generalized formulation of the width function
  here.} \cite{rodriguez_2001}. In general terms, the width function can be
expressed as:

\begin{equation}
  \label{eq:prob1}
  H(t, v_i) = \sum_{\gamma \in \Gamma_i} I(\gamma, t)
\end{equation}

Along with an indicator function, $I(\gamma, t)$:

\begin{equation}
  \label{eq:prob1}
  I(\gamma, t) =
  \begin{cases}
    1 & T(\gamma) = t \\
    0 & \text{otherwise}
  \end{cases}
\end{equation}

Where $\Gamma_i$ is the set of all directed paths to the target vertex $v_i$,
and $T(\gamma)$ is the travel time along path $\gamma$. If the travel times
between vertices are constant, the width function of the graph at vertex $v_i$
can be described as a linear function of the adjacency matrix:\footnote{While
  mathematically concise, this equation is computationally inefficient. See
  Section S1 in the Supplementary Information for the efficient implementation
  used in our analysis.}

\begin{equation}
  H(t, v_i) = (A^t \mathbf{1}) (i)
\end{equation}

In real-world drainage networks, travel times between grid cells are not
uniform. Crucially, the travel time for channelized cells will be roughly 1-2
orders of magnitude faster than the travel time in hillslope cells
\cite{rodriguez_2001, tak_1990}. Thus, to account for this discrepancy, we
define $\phi$ to represent the ratio of hillslope to channel travel times:

\begin{equation}
  \label{eq:prob1}
  \phi = \frac{t_h}{t_c}
\end{equation}

\begin{figure*}[t] \centering
\includegraphics[width=\textwidth]{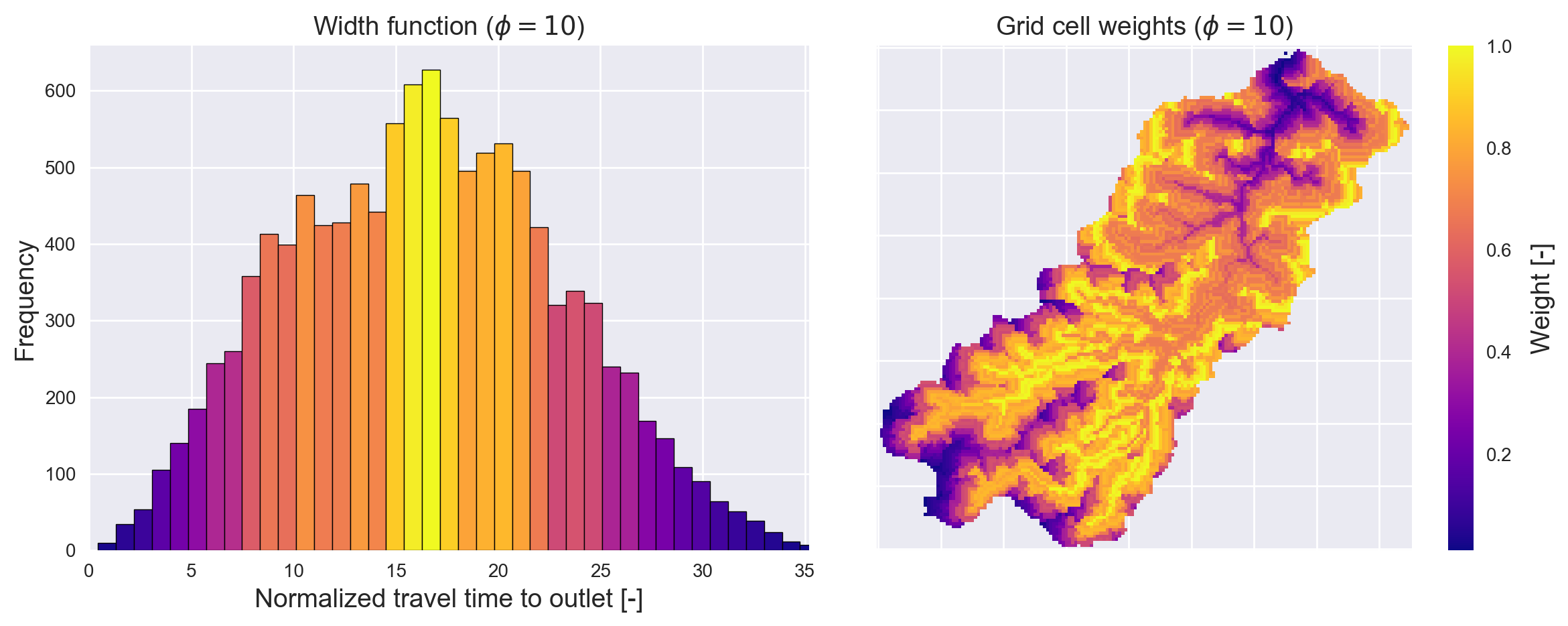}
\caption[Distance histogram and vertex weights]{Left: width function
  (travel-time histogram) of the watershed, assuming that channelized travel
  time is ten times faster than on hillslopes ($\phi = 10$). Right: weights
  associated with each vertex of the graph. Brighter regions correspond to areas
  that contribute to the peaks of the width function.}
  \label{fig:hist_and_ratio}
\end{figure*}

Where $t_h$ is the travel time for hillslopes and $t_c$ is the travel time for
channels. Figure \ref{fig:hist_and_ratio} (left) shows the width function for an
example watershed, under the assumption that channel velocity is ten times
faster than hillslope velocity ($\phi = 10$). The width functions for various
values of $\phi$ are shown in Figures S1 and S2 in the Supplementary Information.

Note that when the effects of hydraulic dispersion are ignored, the width
function is equivalent to the geomorphological impulse unit hydrograph (GIUH) of
the basin \cite{rodriguez_2001}. The GIUH represents the response of the basin
to an instantaneous impulse of rainfall distributed uniformly over the
catchment; or equivalently, the probability that a particle injected randomly
within the watershed at time $t=0$ exits the watershed through the outlet at
time $t=t'$.

\textbf{Accumulation}: The accumulation at vertex $v_i$ describes the number of
vertices located upstream of $v_i$ (or alternatively, the upstream area
\cite{moore_1991}). It is equivalent to the cumulative sum of the width function
with respect to time\footnote{See Section S1 in the Supplementary Information
  for the efficient implementation of the accumulation algorithm.}:

\begin{equation} C(v_i) = (\sum_{t=0}^\infty A^t \mathbf{1}) (i)
\end{equation}

\begin{figure*}[t] \centering
\includegraphics[width=\textwidth]{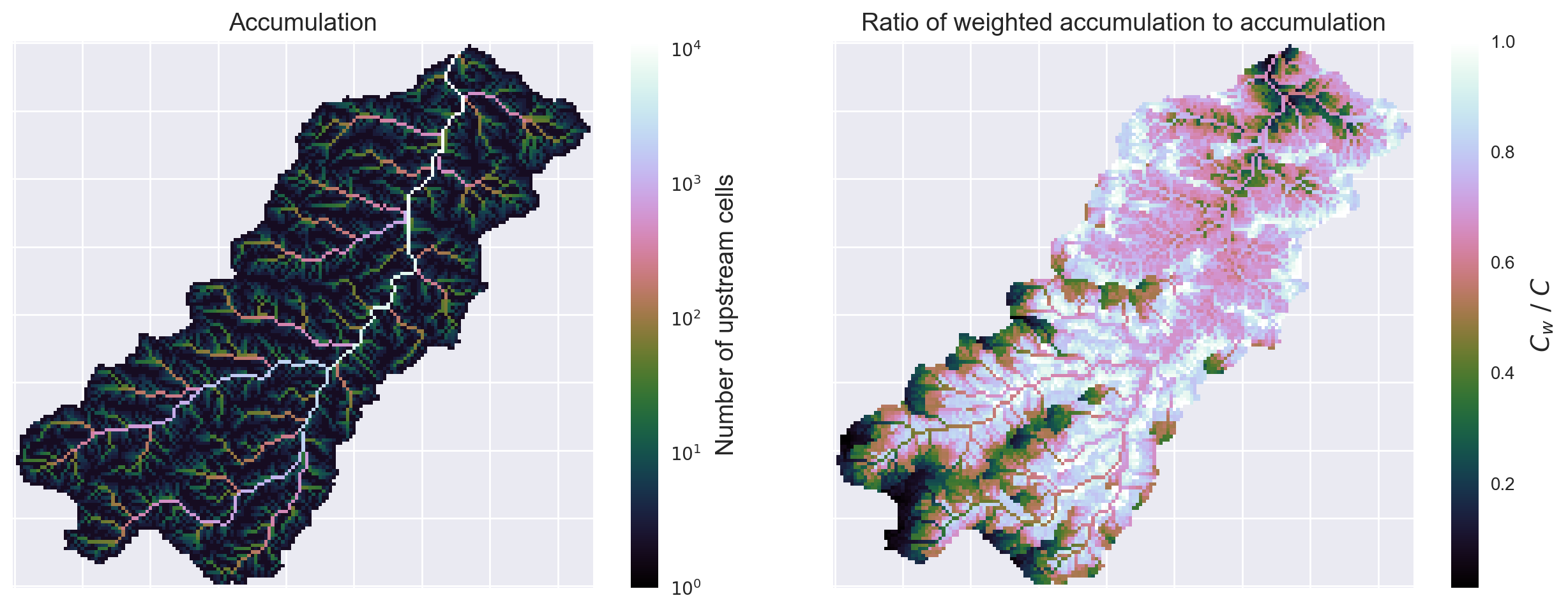}
\caption[Accumulation and weighted accumulation]{Left: accumulation (number of
  cells upstream of every cell). Right: ratio of weighted accumulation to
  accumulation ($C_w / C$).}
  \label{fig:acc_and_wacc}
\end{figure*}

Figure \ref{fig:acc_and_wacc} (left) shows the accumulation at each vertex for
an example catchment. Because upstream area is correlated with mean discharge
\cite{rodriguez_2001}, accumulation is frequently used to determine
locations of channels within a drainage network \cite{moore_1991}.

\textbf{Weighting function}: To identify the vertices that contribute most to
synchronous flows at the outlet, we propose a weighting function that weights
each vertex by its rank in the travel time distribution. Let $\tau_{ij}$
represent the known travel time from a starting vertex $v_j$ to the outlet
vertex $v_i$. Then the weight associated with vertex $v_j$ can be expressed in
terms of a weighting function $W(v_i,v_j)$:

\begin{equation} w_{j} = W(v_i, v_j) = \frac{H(\tau_{ij}, v_i)}{\underset{t}{\max}(H(v_i))}
\end{equation}

Where $\tau_{ij}$ represents the travel time from vertex $v_j$ to vertex $v_i$,
$H(\tau_{ij}, v_i)$ represents the width function for an outlet vertex $v_i$
evaluated at time $\tau_{ij}$, and the normalizing factor
$\underset{t}{\max}(H(v_i))$ represents the maximum value of the width function
over all time steps $t$. In this formulation, vertices are weighted by the rank
of the associated travel time in the width function. Vertices that
contribute to the maximum value of the width function (the mode of the travel
time distribution) will receive the highest possible weight (unity), while
vertices that contribute to the smallest values of the width function will
receive small weights. In other words, vertices will be weighted in proportion
to the number of vertices that share the same travel time to the outlet. Figure
\ref{fig:hist_and_ratio} shows the weights corresponding to each bin of the
travel time distribution (left), along with the weights applied to each vertex
(right). Weights for varying values of $\phi$ are shown in Figures S1 and S2 in
the Supplementary Information.

\textbf{Weighted accumulation}: Much like the \textit{accumulation} describes the number
of vertices upstream of each vertex $v_i$, the \textit{weighted accumulation} yields the
sum of the weights upstream of $v_i$. If each vertex $v_j$ is given a weight
$w_j$, the weighted accumulation at vertex $v_i$ can be defined:

\begin{equation} C_w(v_i, \mathbf{w}) = (\sum_{t=0}^\infty A^t \mathbf{w}) (i)
\end{equation}

Where $\mathbf{w}$ is a vector of weights, with each weight $w_j$ associated
with a vertex $v_j$ in the graph. When the previously-defined weighting function
is used, the weighted accumulation score measures the extent to which a
subcatchment delineated at vertex $v_i$ contributes to synchronous flows at the
outlet. In other words, if the ratio of weighted accumulation to accumulation is
large for a particular vertex, this means that the subcatchment upstream of that
vertex contributes disproportionately to the peak of the hydrograph. Figure
\ref{fig:acc_and_wacc} (right) shows the ratio of weighted accumulation to
accumulation for the example catchment. The weighted accumulation provides a
natural metric for detecting the cuts in the drainage network that will
maximally remove synchronous flows, and thus forms the basis of the controller
placement algorithm.

\subsection{Controller placement algorithm definition}

The controller placement algorithm is described as follows. Let $A$ represent
the adjacency matrix of a watershed delineated at some vertex $v_i$.
Additionally, let $k$ equal the desired number of controllers, and $c$ equal the
maximum upstream accumulation allowed for each controller. The graph is then
partitioned according to the following scheme:

\begin{enumerate}
  \item Compute the width function, $H(t, v_i)$, for the graph described
by adjacency matrix $A$ with an outlet at vertex $v_i$.
  \item Compute the accumulation $C(v_j)$ at each vertex $v_j$.
  \item Use $H(t, v_i)$ to compute the weighted accumulation $C_w(v_j)$ at each vertex $v_j$.
  \item Find the vertex $v_{opt}$, where the accumulation $C(v_{opt})$ is less
than the maximum allowable accumulation and the weighted accumulation
$C_w(v_{opt})$ is maximized:
    \begin{equation} v_{opt} \gets \underset{ v_s \in V_s }{\text{argmax}}(C_{w}(v_s))
    \end{equation}

    Where $V_s$ is the set of vertices such that vertex $v_i$ is reachable from
    any vertex in $V_s$ and the accumulation $C$ at any vertex in $V_s$ is less
    than $c$.
  \item Prune the graph at vertex $v_{opt}$: $A \gets A_p(A, v_{opt})$
  \item If the cumulative number of partitions is equal to $k$, end the
algorithm. Otherwise, start at (1) with the catchment described by the new $A$
matrix.
\end{enumerate}

The algorithm is described formally in Algorithm \ref{alg:1}. An open-source
implementation of the algorithm in the \textit{Python} programming language is
also provided \cite{this_repo_2018}, along with the data needed to reproduce our
results. Efficient implementations of the \textit{delineation},
\textit{accumulation}, and \textit{width function} operations are provided via
the \texttt{pysheds} toolkit, which is maintained by the authors
\cite{pysheds_2018}.

\begin{figure*}[t] \centering
\includegraphics[width=\textwidth]{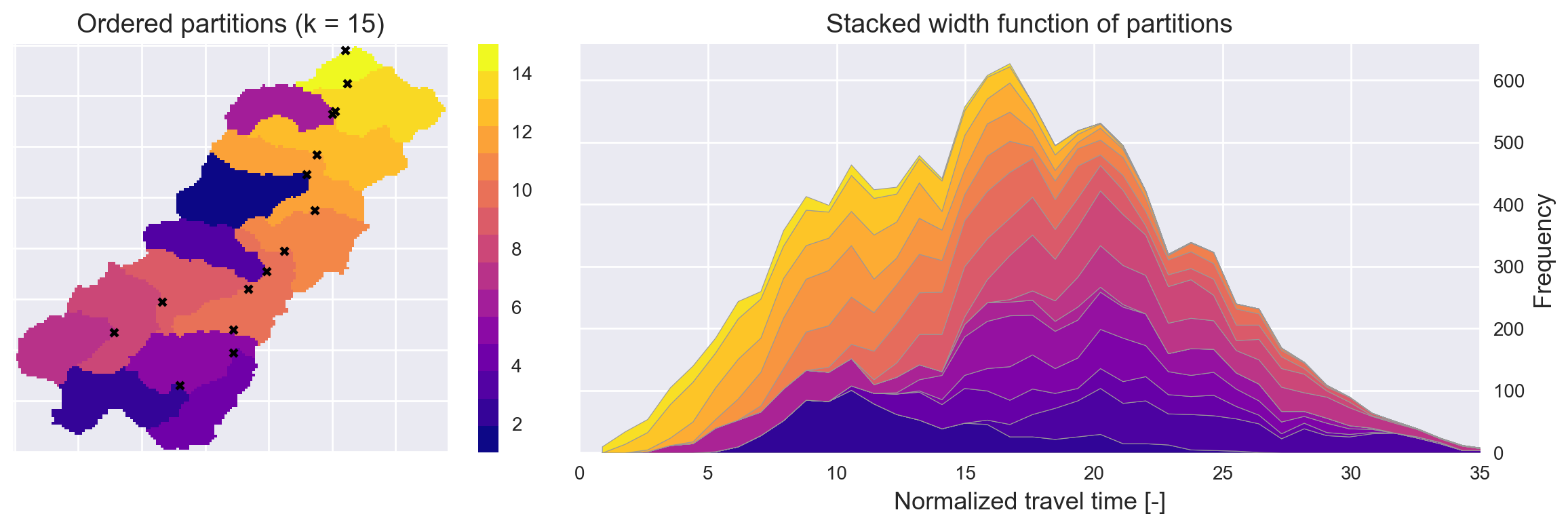}
\caption[Partitions and stacked histogram]{Left: partitioning of the example
  watershed using the controller placement algorithm. Right: stacked width
  functions for each partition. The brightness expresses the priority of each
  partition, with the darker partitions being prioritized over the brighter
  ones.}
  \label{fig:partitions}
\end{figure*}

Figure \ref{fig:partitions} shows the controller configuration generated by
applying the controller placement algorithm to the example watershed, with
$k=15$ controllers, each with a maximum accumulation of $c = 900$ (i.e. each
controller captures roughly 8\% of the catchment's land area). In the left
panel, partitions are shown in order of decreasing priority from dark to light
(i.e. darker regions are partitioned first by the algorithm). The right panel
shows the stacked width functions for each partition. The sum of the width
functions from each partition reconstitute the original width function for the
catchment. From the stacked width functions, it can be seen that the algorithm
tends to prioritize the pruning of subgraphs that align with the peaks of the
travel time distribution. Note for instance, how the least-prioritized paritions
gravitate towards the low end of the travel-time distribution, while the
most-prioritized partitions are centered around the mode. Controller placement
schemes corresponding to different numbers of controllers are shown in Figure S3
in the Supplementary Information.

\begin{algorithm}
\begin{tcolorbox}
 \KwData{\\
 A directed graph described by adjacency matrix $A$\;
 A target vertex $v_i$ with index $i$\;
 A desired number of partitions $k$\;
 The maximum accumulation at each controller, $c$\;}
 \KwResult{Generate partitions for a catchment}
 \vspace{6pt}
 Let $\mathbf{q}$ be a vector representing all vertices in the graph\;
 Let $k_c$ equal the current number of partitions\;
 Let $\mathbf{\tau}$ represent a vector of travel times from each vertex to vertex $v_i$\;
 Let $A$ represent the adjacency matrix of the system\;
 Let $A_c$ represent the adjacency matrix for the current iteration\;
 \vspace{6pt}
 
 $A_c \gets A$\;
 $k_c \gets 0$\;
 
 \While{$k_c < k$}{
 
  $H(t, v_i) \gets (A_c^t \mathbf{1}) (i)$\;
  $C \gets (\sum_{t=0}^\infty A_c^t \mathbf{1})$\;
  $\mathbf{w} \gets W(v_i, \mathbf{q})$\;
  $C_{w} \gets (\sum_{t=0}^\infty A_c^t \mathbf{w})$\;

  \eIf{$C(v_i) > 0$}{
      $V_c \gets \{ v_m \in V | C(v_m) \leq c \}$\;
      $V_s \gets V_d(A_c, v_i) \cap V_c$\;
      $v_{opt} \gets \underset{ v_s \in V_s }{\text{argmax}}(C_{w}(v_s))$\;
      $A_c \gets A_p(A_c, v_{opt})$\;
      $k_c \gets k_c + 1$\;
   }{
  }
 }
 \caption{Controller placement algorithm}
 \label{alg:1}
\end{tcolorbox}
\end{algorithm}

\section{Algorithm validation}

To evaluate the controller placement algorithm, we simulate the controlled
network using a hydrodynamic model, and compare the performance to a series of
randomized controller placement configurations. Performance is characterized by
the ``flatness" of the flow profile at the outlet of the watershed, as measured
by both the peak discharge and the variance of the hydrograph (i.e. the extent
to which the flow deviates from the mean flow over the course of the hydrologic
response). To establish a basis for comparison, we simulate a volume capture
scenario \cite{Emerson_2005}, wherein roughly half of the total contributing
area is controlled, and each controller completely captures the discharge from
its respective upstream area.

The validation experiment is designed to test the central premises of the
controller placement algorithm: that synchronous cells can be identified from
the structure of the drainage network, and that maximally capturing these
synchronous cells will lead to a flatter overall hydrologic response. If these
premises are accurate, we expect to see two results. First, the controller
placement algorithm will produce flatter flows than the randomized control
trials. Second, the performance of the algorithm will be maximized when using a
large number of small partitions. Using many small partitions allows the
algorithm to selectively target the highly-weighted cells that contribute
disproportionately to the peak of the hydrograph. Conversely, large partitions
capture many extraneous low-weight cells that don't contribute to the peak of
the hydrograph. In other words, if increasing the number of partitions improves
the performance of the algorithm, it not only confirms that the algorithm works
for our particular experiment, but also justifies the central premises on
which the algorithm is based.

\subsection{Experimental design}

We evaluate controller configurations based on their ability to flatten the
outlet hydrograph of a test watershed when approximately 50\% of the
contributing area is controlled. This test case is chosen because it presents a
practical scenario with real-world constraints, and because it allows for direct
comparison of many different controller placement strategies. For our test case,
we use the Sycamore Creek watershed, a heavily urbanized creekshed located in
the Dallas--Fort Worth Metroplex with a contributing area of roughly 83 km$^2$
(see Figure \ref{fig:elev_and_river}). This site is the subject of a long-term
monitoring study led by the authors \cite{Bartos_2018}, and is chosen for this
analysis because (i) it is known to experience issues with flash flooding, and
(ii) it is an appropriate size for our analysis---being large enough to capture
fine-scale network topology, but not so large that computation time becomes
burdensome.

A model of the stream network is generated from a conditioned digital elevation
model (DEM) by determining flow directions from the elevation gradient and then
assigning channels to cells that fall above an accumulation threshold.
Conditioned DEMs and flow direction grids at a resolution of 3 arcseconds
(approximately 70 by 90 m) are obtained from the USGS HydroSHEDS database
\cite{Lehner_2008}. Grid cells with an accumulation greater than 100 are defined
to be channelized cells, while those with less than 100 accumulation are defined
as hillslope cells. This threshold is based on visual comparison with the stream
network defined in the National Hydrography Dataset (NHD) \cite{nhd_2013}.
Hillslope cells draining into a common channel are aggregated into
subcatchments, with a flow length corresponding to the longest path within each
hillslope, and a slope corresponding to the average slope over all flow paths in
the subcatchment. To avoid additional complications associated with modeling
infiltration, subcatchments and channels are assumed to be completely
impervious. Channel geometries are assigned to each link within the channelized
portion of the drainage network. We assume that each stream segment can be
represented by a ``wide rectangular channel", which is generally accurate for
natural river reaches in which the stream width is large compared to the stream
depth \cite{mays_2010}. To simulate channel width and depth, we assume a power
law relation between accumulation and channel size based on an empirical
formulation from Moody and Troutman (2002) \cite{Moody_2002}:

\begin{equation}
  \begin{split}
    \omega = 7.2 \cdot Q^{0.50 \pm 0.02} \\
    h = 0.27 \cdot Q^{0.30 \pm 0.01}
    \end{split}
\end{equation}

Where $\omega$ is stream width, $h$ is stream depth, and $Q$ is the mean river
discharge. Knowing the width and depth of the most downstream reach, and
assuming that the accumulation at a vertex is proportional to the mean flow, we
generate channel geometries using the mean parameter values from the above
relations.

Using the controller placement algorithm, control structures are placed such
that approximately 50$\pm$3\% of the catchment area is captured by storage
basins. To investigate the effect of the number of controllers on performance,
optimized controller strategies are generated using between $k=1$ and $k=35$
controllers. The ratio of hillslope-to-channel travel times is assumed to be
$\phi = 10$. We compare the performance of our controller placement algorithm to
randomized controller placement schemes, in which approximately 50$\pm$3\% of
the catchment area is controlled but the placement of controllers is random. For
this comparison assessment, we generate 50 randomized controller placement
trials, each using between $k=1$ and $k=24$ controllers.\footnote{While the
  controller randomization code was programmed to use between 1 and 35
  controllers, the largest number of controllers achieved was 24. This result
  stems from the fact that the randomization algorithm struggled to achieve 30+
  partitions without selecting cells that fell below the channelization
  threshold (100 accumulation).}

We simulate the hydrologic response using a hydrodynamic model, and evaluate
controller placement performance based on the flatness of the resulting
hydrograph. To capture the hydrologic response under various rainfall
conditions, we simulate small, medium and large rainfall events, corresponding
to 0.5, 1.5 and 4.0 mm of rainfall delivered instantaneously over the first five
minutes of the simulation. A hydrodynamic model is used to simulate the
hydrologic response at the outlet by routing runoff through the channel network
using the dynamic wave equations \cite{swmm_2018}. The simulation performance is
measured by both the peak discharge and the total variance of the hydrograph.
The variance of the hydrograph (which we refer to as ``flashiness'') is defined
as:

\begin{equation}
  \label{eq:prob1}
  \sigma^2 = \frac{1}{N} \sum_{i=1}^N (Q_i - \bar{Q})^2 
\end{equation}

Where $Q$ is the discharge, $\bar{Q}$ is the mean discharge in the storm window,
and $N$ is the number of data points in the storm window. This variance metric
captures the flow's deviation from the mean over the course of the hydrologic
response, and thus provides a natural metric for the flatness of the hydrograph.
This metric is important for water quality considerations like first flush
contamination or streambed erosion---in which the volume of transported material
(e.g. contaminants or sediments) depends not only on the maximum discharge, but
also on the duration of flow over a critical threshold \cite{Wong_2016}.

Note that the validation experiment is not intended to faithfully reproduce the
precise hydrologic response of our chosen study area, but rather, to test the
basic premises of the controller placement algorithm. As such, site-specific
details---such as land cover, soil types and existing infrastructure---have been
deliberately simplified. For situations in which these characteristics exert an
important influence on the hydrologic response, one may account for these
factors by adjusting the inter-vertex travel times used in the controller
placement algorithm.

\section{Results}

\begin{figure*}[t]
\centering
\includegraphics[width=\textwidth]{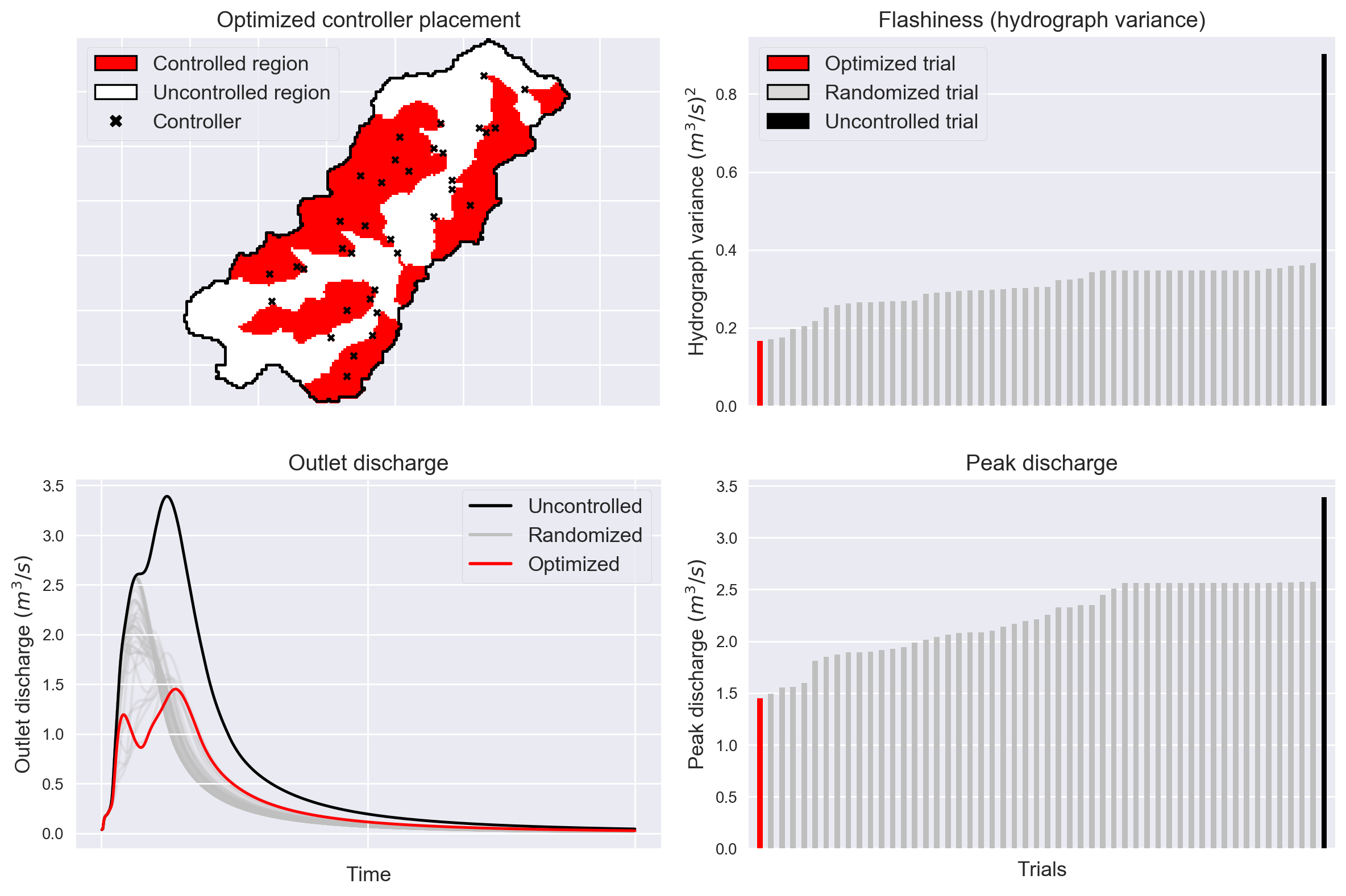}
\caption[Hydraulic modeling results]{Results of the hydraulic simulation
  experiment for the medium storm event (1.5 mm). Top left: optimal controller
  placement ($k=35$), with captured regions in red. Bottom left: hydrographs
  resulting from each simulation. The uncontrolled simulation is shown in black,
  while the optimized controller placement simulations are shown in red, and the
  randomized controller simulations are shown in gray. Right: the overall
  flashiness (variance of the hydrograph) and peak discharge for each
  simulation, using the same coloring scheme.}
\label{fig:swmm}
\end{figure*}

The controller placement algorithm produces consistently flatter flows than
randomized control trials. Figure \ref{fig:swmm} shows the results of the
hydraulic simulation assessment in terms of the resulting hydrographs (bottom
left), and the overall flashiness and peak discharge of each simulation (right)
for the medium-sized (1.5 mm) storm event. The best performance is achieved by
using the controller placement algorithm with $k=35$ controllers (see Figure
\ref{fig:swmm}, top left). Comparing the overall variances and peak discharges,
it can be seen that the optimized controller placement produces flatter outlet
discharges than any of the randomized controller placement strategies.
Specifically, the optimized controller placement achieves a peak discharge that
is roughly 47\% of that of the uncontrolled case, while the randomized
simulations by comparison achieve an average peak discharge that is more than
72\% of that of the uncontrolled case. Similarly, the hydrograph variance of the
optimized controller placement is roughly 21\% of that of the uncontrolled case,
compared to 35\% for the randomized simulations on average.\footnote{Note that
  the controller placement algorithm results in a longer falling limb than the
  randomized trials. This result stems from the fact that the algorithm
  prioritizes the removal of grid cells that contribute to the peak and rising
  limb of the hydrograph, while grid cells contributing to the falling limb are
  ignored. In other words, the controller placement algorithm shifts discharges
  from the peak of the hydrograph to the falling limb.} When tested against
storm events of different sizes (0.5 to 4 mm of rain), the controller placement
algorithm also generally outperforms randomized control trials (see Section S4
in the Supplementary Information). However, the within-group performance varies
slightly with rain event size, which could result from the nonlinearities
inherent in wave propagation speed. Thus, while the optimized controller
placement still produces flatter flows than randomized controls, this result
suggests that the performance of the controller placement algorithm could be
further improved by tuning the assumed inter-vertex travel times to correspond
to the expected speed of wave propagation.

\begin{figure*}[t]
\centering
\includegraphics[width=\textwidth]{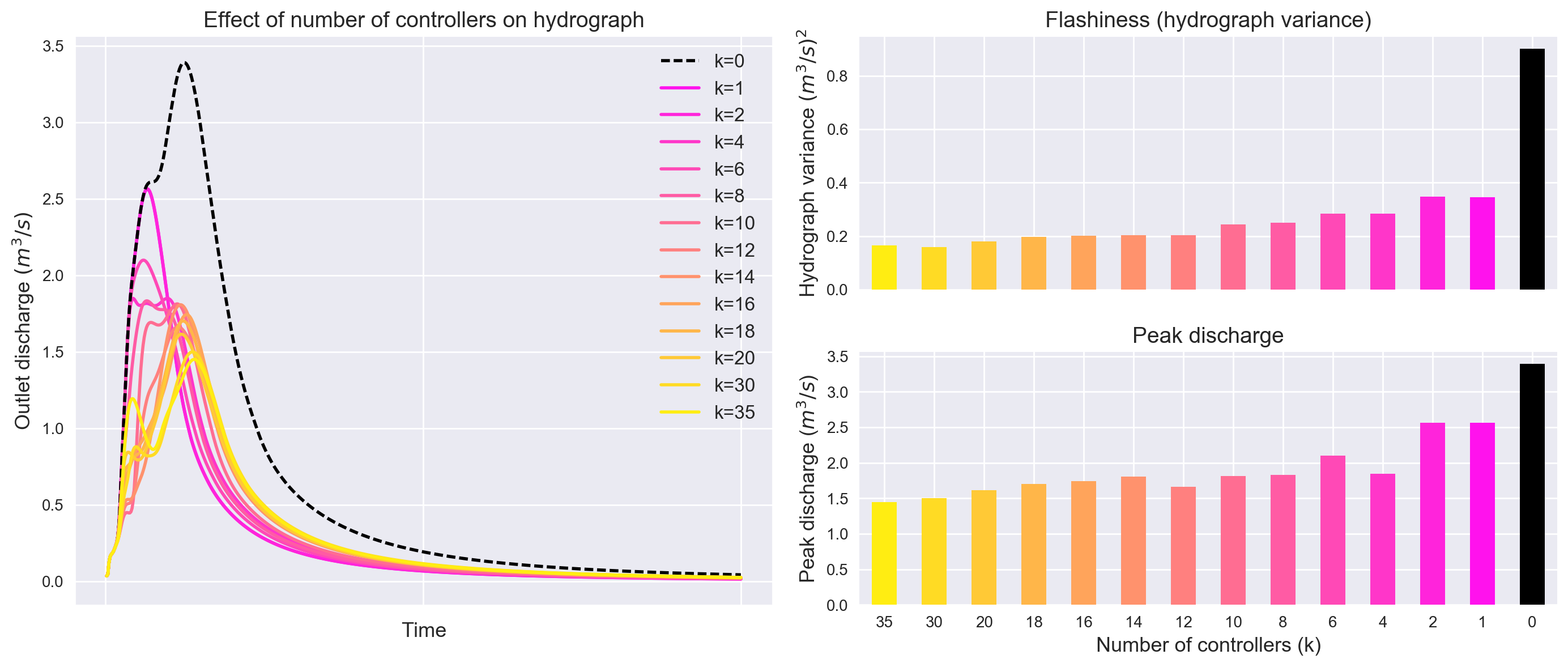}
\caption{Left: hydrographs associated with varying numbers of controllers (k),
  using the controller placement algorithm with 50\% watershed area
  removal. Right: hydrograph variance (top) and peak discharge (bottom) vs.
  number of controllers. In general, more controllers produces a flatter response.}
\label{fig:num_controllers}
\end{figure*}

Under the controller placement algorithm, the best performance is achieved by
using a large number of small-scale controllers; however, more controllers does
not lead to better performance for the randomized controller placement scheme.
Given that increasing the number of controllers allows the algorithm to better
target highly synchronous cells, this result is consistent with the central
premise that capturing synchronous cells will lead to a flatter hydrologic
response. Figure \ref{fig:num_controllers} shows the optimized hydrologic
response for varying numbers of controllers (left), along with the overall
variance (top right) and peak discharge (bottom right). In all cases, roughly
50\% of the watershed is controlled; however, configurations using many small
controllers consistently perform better than configurations using a few large
controllers. This trend does not hold for the randomized controller placement
strategy (see Figure S8 in the Supplementary Information). Indeed, the
worst-performing randomized controller placement uses $k=6$ controllers (out of
a minimum of 1) while the best-performing randomized controller placement uses
$k=18$ controllers (out of a maximum of 24). The finding that the controller
placement algorithm converges to a (locally) optimal solution follows from the
fact that as the number of partitions increases, controllers are better able to
capture highly-weighted regions without also capturing extraneous low-weight
cells. This in turn implies that the weighting scheme used by the algorithm
accurately identifies the regions of the watershed that contribute
disproportionately to synchronized flows. Thus, in spite of various sources of
model and parameter uncertainty, the experimental results confirm the central
principles under which the controller placement algorithm operates: namely, that
synchronous regions can be deduced from the graph structure alone, and that
controlling these regions results in a flatter hydrograph compared to randomized
controls.

\begin{figure*}[t]
\centering
\includegraphics[width=\textwidth]{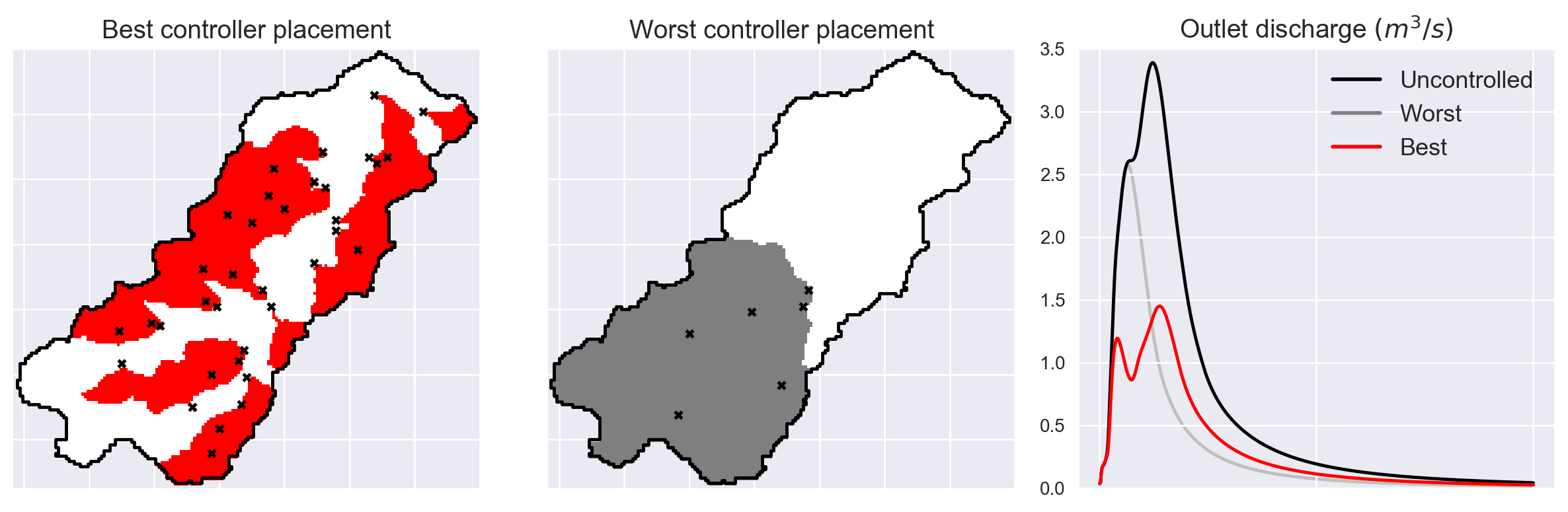}
\caption{Left: Best controller placement in terms of peak discharge ($k = 35$
  controllers, optimized). Center: worst controller placement in terms of peak
  discharge ($k = 6$ controllers, randomized). Controller locations are
  indicated by black crosses, and controlled partitions are indicated by colored
  regions. Right: hydrographs associated with the best and worst controller
  placement strategies.}
\label{fig:placement_visualization}
\end{figure*}

In addition to demonstrating the efficacy of the controller placement algorithm,
the validation experiments reveal some general principles for organizing
hydraulic control structures within drainage networks to achieve downstream
streamflow objectives. Overall, the controller placement strategies that perform
best---whether achieved through optimization or randomization---tend to
partition the watershed axially rather than laterally. These lengthwise
partitions result in a long, thin drainage network that prevents tributary flows
from ``piling up''. Figure \ref{fig:placement_visualization} shows the
partitions corresponding to the best-performing and worst-performing controller
placement strategies with respect to peak discharge (left and center,
respectively), along with the associated hydrographs (right). While the
best-performing controller placement strategy evenly distributes the partitions
along the length of the watershed, the worst-performing controller placement
strategy controls only the most upstream half of the watershed. As a result, the
worst-performing strategy removes the largest part of the peak, but completely
misses the portion of the peak originating from the downstream half of the
watershed. In order to achieve a flat downstream hydrograph, controller
placement strategies should seek to evenly distribute controllers along the
length of the watershed.

\section{Discussion}

The controller placement algorithm presented in this study provides a tool for
designing stormwater control systems to better mitigate floods, regulate
contaminants, and protect aquatic ecosystems. By reducing peak discharge,
optimized placement of stormwater control structures may help to lessen the
impact of flash floods. Existing flood control measures often focus on
controlling large riverine floods---typically through existing centralized
assets, like dams and levees. However, flash floods may occur in small
tributaries, canals, and even normally dry areas. For this reason, flash floods
are not typically addressed by large-scale flood control measures, despite the
fact that they cause more fatalities than riverine floods in the developed world
\cite{Doocy_2013}. By facilitating distributed control of urban flash floods,
our controller placement strategy could help reduce flash flood related
mortality. Moreover, by flattening the hydrologic response, our controller
placement algorithm promises to deliver a number of environmental and water
quality benefits, such as decreased first flush contamination \cite{Wong_2016},
decreased sediment transport \cite{muschalla_2014}, improved potential for
treatment in downstream green infrastructure \cite{Kerkez_2016, Bartos_2018},
and regulation of flows in sensitive aquatic ecosystems \cite{poresky_2015}.

\subsection{Key features of the algorithm}

The controller placement algorithm satisfies a number of important operational
considerations:

\begin{itemize}
\item \textbf{Theoretically motivated}. The controller placement algorithm has
  its foundation in the theory of the geomorphological impulse unit
  hydrograph---a relatively mature theory supported by a established body of
  research \cite{rodriguez_2001, kirkby_1976, gupta_1986, gupta_1988, mesa_1986,
    marani_1991, troutman_1985}. Moreover, the algorithm works in an intuitive
  way---by recursively removing the subcatchments of a watershed that contribute
  most to synchronized flows. This theoretical basis distinguishes our algorithm
  from other strategies that involve exhaustive optimization or direct
  application of existing graph theoretical constructs (such as graph centrality
  metrics).

\item \textbf{Generalizable and extensible}. Because it relies solely on network
  topology, the controller placement algorithm will provide consistent results
  for any drainage network---including both natural stream networks and
  constructed sewer systems. Moreover, because each step in our algorithm has a
  clear meaning in terms of the underlying hydrology, the algorithm can be
  modified to satisfy more complex control problems (such as systems in which
  specific regulatory requirements must be met).

 \item \textbf{Flexible to user objectives and constraints}. The controller
   placement algorithm permits specification of important practical constraints,
   such as the amount of drainage area that each control site can capture, and
   the number of control sites available. Moreover, the weighting function can
   be adjusted to optimize for a variety of objectives (such as the overall
   ``flatness'' of the hydrograph, or removal of flows from a contaminated
   upstream region).
  
 \item \textbf{Parsimonious with respect to data requirements}. The controller
   placement algorithm requires only a digital elevation model of the watershed
   of interest. Additional data---such as land cover and existing hydraulic
   infrastructure---can be used to fine-tune estimates of travel times within
   the drainage network, but are not required by the algorithm itself.

 \item \textbf{Fast implementation} For the watershed examined in this study
   (consisting of about 12,000 vertices), the controller placement algorithm
   computes optimal locations for $k=15$ controllers in roughly
   3.0 seconds (on a 2.9 GHz Intel Core i5 processor). While the
   computational complexity of the algorithm is difficult to
   characterize\footnote{The computational complexity of the controller
     placement algorithm depends on the implementation of component functions
     (such as delineation and accumulation computation), which can in turn
     depend on the structure of the watershed itself.}, it is faster than other
   comparable graph-cutting algorithms, such as recursive spectral bisection or
   spectral clustering, both of which are $O(n^3)$ in computational complexity.
\end{itemize}

Taken together, our algorithm offers a solution to the controller placement
problem that is suitable for research as well as for practical applications. On
one hand, the algorithm is based in hydrologic and geomorphological theory, and
provides important insights into the connections between geomorphology and the
design of the built environment. On the other hand, the algorithm is fast,
robust, and easy-to-use, making it a useful tool for practicing engineers and
water resource managers.

\subsection{Caveats and directions for future research}

While our controller placement algorithm is robust and broadly-applicable, there
are a number of important considerations to keep in mind when applying this
algorithm to real-world problems.

\begin{itemize}
\item The controller placement algorithm implicitly assumes that rainfall is
  uniform over the catchment of interest. While this assumption is justified for
  small catchments in which the average spatial distribution of rainfall will be
  roughly uniform, this assumption may not hold for large (e.g. continent-scale)
  watersheds. Modifications to the algorithm would be necessary to account for a
  non-uniform spatial distribution of rainfall.

\item The controller placement algorithm is sensitive to the chosen ratio of
  hillslope to channel speeds, $\phi$. Care should be taken to select an
  appropriate value of $\phi$ based on site-specific land cover and
  morphological characteristics. More generally, for situations in which
  differential land cover, soil types, and existing hydraulic infrastructure
  play a dominating role, the performance of the algorithm may be enhanced by
  adjusting inter-vertex travel times to correspond to estimated overland flow
  and channel velocities.

\item Our assessment of the algorithm's performance rests on the assumption that
  installed control structures (e.g. retention basins) are large enough to
  capture upstream discharges. The algorithm itself does not explicitly account
  for endogenous upstream flooding that could be introduced by installing new
  control sites.

\item In this study, experiments were conducted only for impulsive rainfall
  inputs (i.e. with a short duration of rainfall). Future work should assess the
  performance of the distance-weighted controller placement strategy under
  arbitrary rainfall durations.

\end{itemize}

More broadly, future research should investigate the problem of sensor placement
in stream networks using the theoretical framework developed in this paper.
While this study focuses on the problem of optimal placement of hydraulic
control structures, our algorithm also suggests a solution to the problem of
sensor placement. Stated probabilistically, the geomorphological impulse unit
hydrograph (GIUH) represents the probability that a ``particle'' injected
randomly within the watershed at time $t=0$ exits the outlet at time $t=t'$. Thus,
the peaks of the GIUH correspond to the portions of the hydrologic response
where there is the greatest amount of ambiguity about where a given ``particle''
originated. It follows that the same locations that maximally de-synchronize
flows may also be the best locations for disambiguating the locations from which
synchronous flows originated. Future experiments should investigate the ability
to estimate upstream states (e.g. flows) within the network given an outlet
discharge along with internal state observers (e.g. flow sensors) placed using
the algorithm developed in this study.

\section{Conclusions}

We develop an algorithm for placement of hydraulic control structures that
maximally flattens the hydrologic response of drainage networks. This algorithm
uses the geomorphological impulse unit hydrograph to locate subcatchments that
dominate the peaks of the hydrograph, then partitions the drainage network to
minimize the contribution of these subcatchments. We find that the controller
placement algorithm produces flatter hydrographs than randomized controller
placement trials---both in terms of peak discharge and overall variance. By
reducing the flashiness of the hydrologic response, our controller placement
algorithm may one day help to mitigate flash floods and restore urban water
quality through reduction of contaminant loads and prevention of streambed
erosion. We find that the performance of the algorithm is enhanced when using a
large number of small, distributed controllers. In addition to confirming the
central hypothesis that synchronous cells can be identified based on network
structure of drainage basins, this result lends justification to the development
of decentralized \textit{smart} stormwater systems, in which active control of
small-scale retention basins, canals and culverts enables more effective
management of urban stormwater. Overall, our algorithm is efficient, requires
only digital elevation model data, and is robust to parameter and model
uncertainty, making it suitable both as a research tool, and as a design tool
for practicing water resources engineers.

\section*{Acknowledgments}
Funding for this project was provided by the National Science Foundation (Grants
1639640 and 1442735) and the University of Michigan. We would like to thank Alex
Ritchie for exploring alternative approaches to the controller placement problem
and for his help with literature review. We would also like to thank Dr. Alfred
Hero for his advice in formulating the problem.

\section*{Declarations of interest}
Declarations of interest: none.

\section*{Data availability}
Upon publication, code and data links will be made available at:

\url{https://github.com/kLabUM/hydraulic-controller-placement}

\section*{References}

\bibliography{bibliography}

\begin{thebibliography}{10}
\expandafter\ifx\csname url\endcsname\relax
  \def\url#1{\texttt{#1}}\fi
\expandafter\ifx\csname urlprefix\endcsname\relax\def\urlprefix{URL }\fi
\expandafter\ifx\csname href\endcsname\relax
  \def\href#1#2{#2} \def\path#1{#1}\fi

\bibitem{Kerkez_2016}
B.~Kerkez, C.~Gruden, M.~Lewis, L.~Montestruque, M.~Quigley, B.~Wong, A.~Bedig,
  R.~Kertesz, T.~Braun, O.~Cadwalader, A.~Poresky, C.~Pak, Smarter stormwater
  systems, Environmental Science \& Technology 50~(14) (2016) 7267--7273.
\newblock \href {http://dx.doi.org/10.1021/acs.est.5b05870}
  {\path{doi:10.1021/acs.est.5b05870}}.

\bibitem{Doocy_2013}
S.~Doocy, A.~Daniels, S.~Murray, T.~D. Kirsch, The human impact of floods: a
  historical review of events 1980-2009 and systematic literature review,
  {PLoS} Currents{. }\href
  {http://dx.doi.org/10.1371/currents.dis.f4deb457904936b07c09daa98ee8171a}
  {\path{doi:10.1371/currents.dis.f4deb457904936b07c09daa98ee8171a}}.

\bibitem{Wright_2017}
J.~Wright, D.~Marchese, Briefing: Continuous monitoring and adaptive control:
  the `smart' storm water management solution, Proceedings of the Institution
  of Civil Engineers - Smart Infrastructure and Construction 170~(4) (2017)
  86--89.
\newblock \href {http://dx.doi.org/10.1680/jsmic.17.00017}
  {\path{doi:10.1680/jsmic.17.00017}}.

\bibitem{walsh_2005}
C.~J. Walsh, A.~H. Roy, J.~W. Feminella, P.~D. Cottingham, P.~M. Groffman,
  R.~P. Morgan, The urban stream syndrome: current knowledge and the search for
  a cure, Journal of the North American Benthological Society 24~(3) (2005)
  706--723.
\newblock \href {http://dx.doi.org/10.1899/04-028.1}
  {\path{doi:10.1899/04-028.1}}.

\bibitem{booth_1997}
D.~B. Booth, C.~R. Jackson, Urbanization of aquatic systems: degradation
  thresholds, stormwater detection, and the limits of mitigation, Journal of
  the American Water Resources Association 33~(5) (1997) 1077--1090.
\newblock \href {http://dx.doi.org/10.1111/j.1752-1688.1997.tb04126.x}
  {\path{doi:10.1111/j.1752-1688.1997.tb04126.x}}.

\bibitem{finkenbine_2000}
J.~K. Finkenbine, J.~Atwater, D.~Mavinic, Stream health after urbanization,
  Journal of the American Water Resources Association 36~(5) (2000) 1149--1160.

\bibitem{wang_2001}
L.~Wang, J.~Lyons, P.~Kanehl, R.~Bannerman, Impacts of urbanization on stream
  habitat and fish across multiple spatial scales, Environmental Management
  28~(2) (2001) 255--266.
\newblock \href {http://dx.doi.org/10.1007/s0026702409}
  {\path{doi:10.1007/s0026702409}}.

\bibitem{sahagun_2013}
L.~Sahagun, High cost of fighting urban runoff examined in report, LA Times.

\bibitem{wines_2014}
M.~Wines, Behind toledo{'}s water crisis, a long-troubled {Lake} {Erie}, New
  York Times 4.

\bibitem{rosenberg_2010}
E.~A. Rosenberg, P.~W. Keys, D.~B. Booth, D.~Hartley, J.~Burkey, A.~C.
  Steinemann, D.~P. Lettenmaier, Precipitation extremes and the impacts of
  climate change on stormwater infrastructure in washington state, Climatic
  Change 102~(1-2) (2010) 319--349.
\newblock \href {http://dx.doi.org/10.1007/s10584-010-9847-0}
  {\path{doi:10.1007/s10584-010-9847-0}}.

\bibitem{dams_and_development_2000}
Dams and Development: A New Framework for Decision-making - The Report of the
  World Commission on Dams, Routledge, 2000.
\newblock \href {http://dx.doi.org/10.4324/9781315541518}
  {\path{doi:10.4324/9781315541518}}.

\bibitem{arora_2015}
M.~Arora, H.~Malano, B.~Davidson, R.~Nelson, B.~George, Interactions between
  centralized and decentralized water systems in urban context: A review, Wiley
  Interdisciplinary Reviews: Water 2~(6) (2015) 623--634.
\newblock \href {http://dx.doi.org/10.1002/wat2.1099}
  {\path{doi:10.1002/wat2.1099}}.

\bibitem{Hawley_2016}
R.~J. Hawley, G.~J. Vietz, Addressing the urban stream disturbance regime,
  Freshwater Science 35~(1) (2016) 278--292.
\newblock \href {http://dx.doi.org/10.1086/684647} {\path{doi:10.1086/684647}}.

\bibitem{coffman_1999}
L.~S. Coffman, R.~Goo, R.~Frederick, Low-impact development: An innovative
  alternative approach to stormwater management, in: {WRPMD} 1999, American
  Society of Civil Engineers, 1999.
\newblock \href {http://dx.doi.org/10.1061/40430(1999)118}
  {\path{doi:10.1061/40430(1999)118}}.

\bibitem{strecker_2000}
E.~Strecker, M.~M. Quigley, B.~R. Urbonas, J.~Jones, J.~Clary, Determining
  urban stormwater {BMP} effectiveness, Proceedings of the Water Environment
  Federation 2000~(6) (2000) 2395--2412.
\newblock \href {http://dx.doi.org/10.2175/193864700785150457}
  {\path{doi:10.2175/193864700785150457}}.

\bibitem{askarizadeh_2015}
A.~Askarizadeh, M.~A. Rippy, T.~D. Fletcher, D.~L. Feldman, J.~Peng, P.~Bowler,
  A.~S. Mehring, B.~K. Winfrey, J.~A. Vrugt, A.~AghaKouchak, S.~C. Jiang, B.~F.
  Sanders, L.~A. Levin, S.~Taylor, S.~B. Grant, From rain tanks to catchments:
  Use of low-impact development to address hydrologic symptoms of the urban
  stream syndrome, Environmental Science {\&} Technology 49~(19) (2015)
  11264--11280.
\newblock \href {http://dx.doi.org/10.1021/acs.est.5b01635}
  {\path{doi:10.1021/acs.est.5b01635}}.

\bibitem{Bartos_2018}
M.~Bartos, B.~Wong, B.~Kerkez, Open storm: a complete framework for sensing and
  control of urban watersheds, Environmental Science: Water Research {\&}
  Technology 4~(3) (2018) 346--358.
\newblock \href {http://dx.doi.org/10.1039/c7ew00374a}
  {\path{doi:10.1039/c7ew00374a}}.

\bibitem{Mullapudi_2017}
A.~Mullapudi, B.~P. Wong, B.~Kerkez, Emerging investigators series: building a
  theory for smart stormwater systems, Environ. Sci.: Water Res. Technol. 3~(1)
  (2017) 66--77.
\newblock \href {http://dx.doi.org/10.1039/c6ew00211k}
  {\path{doi:10.1039/c6ew00211k}}.

\bibitem{Mullapudi_2018}
A.~Mullapudi, M.~Bartos, B.~Wong, B.~Kerkez, Shaping streamflow using a
  real-time stormwater control network, Sensors 18~(7) (2018) 2259.
\newblock \href {http://dx.doi.org/10.3390/s18072259}
  {\path{doi:10.3390/s18072259}}.

\bibitem{Montestruque_2015}
L.~Montestruque, M.~D. Lemmon, Globally coordinated distributed storm water
  management system, in: Proceedings of the 1st {ACM} International Workshop on
  Cyber-Physical Systems for Smart Water Networks - {CySWater} 2015, {ACM}
  Press, 2015.
\newblock \href {http://dx.doi.org/10.1145/2738935.2738948}
  {\path{doi:10.1145/2738935.2738948}}.

\bibitem{Emerson_2005}
C.~H. Emerson, C.~Welty, R.~G. Traver, Watershed-scale evaluation of a system
  of storm water detention basins, Journal of Hydrologic Engineering 10~(3)
  (2005) 237--242.
\newblock \href {http://dx.doi.org/10.1061/(asce)1084-0699(2005)10:3(237)}
  {\path{doi:10.1061/(asce)1084-0699(2005)10:3(237)}}.

\bibitem{petrucci_2013}
G.~Petrucci, E.~Rioust, J.-F. Deroubaix, B.~Tassin, Do stormwater source
  control policies deliver the right hydrologic outcomes?, Journal of Hydrology
  485 (2013) 188--200.
\newblock \href {http://dx.doi.org/10.1016/j.jhydrol.2012.06.018}
  {\path{doi:10.1016/j.jhydrol.2012.06.018}}.

\bibitem{wong_2018}
B.~P. Wong, B.~Kerkez, Real-time control of urban headwater catchments through
  linear feedback: performance, analysis and site selection, Water Resources
  Research{. }\href {http://dx.doi.org/10.1029/2018wr022657}
  {\path{doi:10.1029/2018wr022657}}.

\bibitem{gelormino_1994}
M.~S. Gelormino, N.~L. Ricker, Model-predictive control of a combined sewer
  system, International Journal of Control 59~(3) (1994) 793--816.
\newblock \href {http://dx.doi.org/10.1080/00207179408923105}
  {\path{doi:10.1080/00207179408923105}}.

\bibitem{mollerup_2016}
A.~L. Mollerup, P.~S. Mikkelsen, D.~Thornberg, G.~Sin, Controlling sewer
  systems - a critical review based on systems in three {EU} cities, Urban
  Water Journal 14~(4) (2016) 435--442.
\newblock \href {http://dx.doi.org/10.1080/1573062x.2016.1148183}
  {\path{doi:10.1080/1573062x.2016.1148183}}.

\bibitem{kirkby_1976}
M.~J. Kirkby, Tests of the random network model, and its application to basin
  hydrology, Earth Surface Processes 1~(3) (1976) 197--212.
\newblock \href {http://dx.doi.org/10.1002/esp.3290010302}
  {\path{doi:10.1002/esp.3290010302}}.

\bibitem{gupta_1986}
V.~K. Gupta, E.~Waymire, I.~Rodr{\'{\i}}guez-Iturbe, On scales, gravity and
  network structure in basin runoff, in: Scale Problems in Hydrology, Springer
  Netherlands, 1986, pp. 159--184.
\newblock \href {http://dx.doi.org/10.1007/978-94-009-4678-1_8}
  {\path{doi:10.1007/978-94-009-4678-1_8}}.

\bibitem{gupta_1988}
V.~K. Gupta, O.~J. Mesa, Runoff generation and hydrologic response via channel
  network geomorphology {\textemdash} recent progress and open problems,
  Journal of Hydrology 102~(1-4) (1988) 3--28.
\newblock \href {http://dx.doi.org/10.1016/0022-1694(88)90089-3}
  {\path{doi:10.1016/0022-1694(88)90089-3}}.

\bibitem{mesa_1986}
O.~J. Mesa, E.~R. Mifflin, On the relative role of hillslope and network
  geometry in hydrologic response, in: Scale Problems in Hydrology, Springer
  Netherlands, 1986, pp. 1--17.
\newblock \href {http://dx.doi.org/10.1007/978-94-009-4678-1_1}
  {\path{doi:10.1007/978-94-009-4678-1_1}}.

\bibitem{marani_1991}
A.~Marani, R.~Rigon, A.~Rinaldo, A note on fractal channel networks, Water
  Resources Research 27~(12) (1991) 3041--3049.
\newblock \href {http://dx.doi.org/10.1029/91wr02077}
  {\path{doi:10.1029/91wr02077}}.

\bibitem{troutman_1985}
B.~M. Troutman, M.~R. Karlinger, Unit hydrograph approximations assuming linear
  flow through topologically random channel networks, Water Resources Research
  21~(5) (1985) 743--754.
\newblock \href {http://dx.doi.org/10.1029/wr021i005p00743}
  {\path{doi:10.1029/wr021i005p00743}}.

\bibitem{Mantilla_2011}
R.~Mantilla, V.~K. Gupta, B.~M. Troutman, Scaling of peak flows with constant
  flow velocity in random self-similar networks, Nonlinear Processes in
  Geophysics 18~(4) (2011) 489--502.
\newblock \href {http://dx.doi.org/10.5194/npg-18-489-2011}
  {\path{doi:10.5194/npg-18-489-2011}}.

\bibitem{Tejedor_2015a}
A.~Tejedor, A.~Longjas, I.~Zaliapin, E.~Foufoula-Georgiou, Delta channel
  networks: 1. a graph-theoretic approach for studying connectivity and steady
  state transport on deltaic surfaces, Water Resources Research 51~(6) (2015)
  3998--4018.
\newblock \href {http://dx.doi.org/10.1002/2014wr016577}
  {\path{doi:10.1002/2014wr016577}}.

\bibitem{Tejedor_2015b}
A.~Tejedor, A.~Longjas, I.~Zaliapin, E.~Foufoula-Georgiou, Delta channel
  networks: 2. metrics of topologic and dynamic complexity for delta
  comparison, physical inference, and vulnerability assessment, Water Resources
  Research 51~(6) (2015) 4019--4045.
\newblock \href {http://dx.doi.org/10.1002/2014wr016604}
  {\path{doi:10.1002/2014wr016604}}.

\bibitem{Zellner_2016}
M.~Zellner, D.~Massey, E.~Minor, M.~Gonzalez-Meler, Exploring the effects of
  green infrastructure placement on neighborhood-level flooding via spatially
  explicit simulations, Computers, Environment and Urban Systems 59 (2016) 116
  -- 128.
\newblock \href {http://dx.doi.org/10.1016/j.compenvurbsys.2016.04.008}
  {\path{doi:10.1016/j.compenvurbsys.2016.04.008}}.

\bibitem{schubert_2017}
J.~E. Schubert, M.~J. Burns, T.~D. Fletcher, B.~F. Sanders, A framework for the
  case-specific assessment of green infrastructure in mitigating urban flood
  hazards, Advances in Water Resources 108 (2017) 55 -- 68.
\newblock \href {http://dx.doi.org/10.1016/j.advwatres.2017.07.009}
  {\path{doi:10.1016/j.advwatres.2017.07.009}}.

\bibitem{yao_2015}
L.~Yao, L.~Chen, W.~Wei, R.~Sun, Potential reduction in urban runoff by green
  spaces in beijing: A scenario analysis, Urban Forestry {\&} Urban Greening
  14~(2) (2015) 300 -- 308.
\newblock \href {http://dx.doi.org/10.1016/j.ufug.2015.02.014}
  {\path{doi:10.1016/j.ufug.2015.02.014}}.

\bibitem{zhang_2015}
B.~Zhang, G.~di~Xie, N.~Li, S.~Wang, Effect of urban green space changes on the
  role of rainwater runoff reduction in beijing, china, Landscape and Urban
  Planning 140 (2015) 8 -- 16.
\newblock \href {http://dx.doi.org/10.1016/j.landurbplan.2015.03.014}
  {\path{doi:10.1016/j.landurbplan.2015.03.014}}.

\bibitem{norton_2015}
B.~A. Norton, A.~M. Coutts, S.~J. Livesley, R.~J. Harris, A.~M. Hunter, N.~S.
  Williams, Planning for cooler cities: A framework to prioritise green
  infrastructure to mitigate high temperatures in urban landscapes, Landscape
  and Urban Planning 134 (2015) 127 -- 138.
\newblock \href {http://dx.doi.org/10.1016/j.landurbplan.2014.10.018}
  {\path{doi:10.1016/j.landurbplan.2014.10.018}}.

\bibitem{meerow_2017}
S.~Meerow, J.~P. Newell, Spatial planning for multifunctional green
  infrastructure: Growing resilience in detroit, Landscape and Urban Planning
  159 (2017) 62 -- 75.
\newblock \href {http://dx.doi.org/10.1016/j.landurbplan.2016.10.005}
  {\path{doi:10.1016/j.landurbplan.2016.10.005}}.

\bibitem{schilling_2008}
J.~Schilling, J.~Logan, Greening the rust belt: A green infrastructure model
  for right sizing america's shrinking cities, Journal of the American Planning
  Association 74~(4) (2008) 451--466.

\bibitem{cattafi_2011}
M.~Cattafi, M.~Gavanelli, M.~Nonato, S.~Alvisi, M.~Franchini, Optimal placement
  of valves in a water distribution network with clp(fd), Theory and Practice
  of Logic Programming 11~(4-5) (2011) 731–747.
\newblock \href {http://dx.doi.org/10.1017/S1471068411000275}
  {\path{doi:10.1017/S1471068411000275}}.

\bibitem{creaco_2010}
E.~Creaco, M.~Franchini, S.~Alvisi, Optimal placement of isolation valves in
  water distribution systems based on valve cost and weighted average demand
  shortfall, Water Resources Management 24~(15) (2010) 4317--4338.
\newblock \href {http://dx.doi.org/10.1007/s11269-010-9661-5}
  {\path{doi:10.1007/s11269-010-9661-5}}.

\bibitem{Perelman_2013}
L.~Perelman, A.~Ostfeld, Application of graph theory to sensor placement in
  water distribution systems, in: World Environmental and Water Resources
  Congress 2013, American Society of Civil Engineers, 2013.
\newblock \href {http://dx.doi.org/10.1061/9780784412947.060}
  {\path{doi:10.1061/9780784412947.060}}.

\bibitem{Yazdani_2011}
A.~Yazdani, P.~Jeffrey, Robustness and vulnerability analysis of water
  distribution networks using graph theoretic and complex network principles,
  in: Water Distribution Systems Analysis 2010, American Society of Civil
  Engineers, 2011.
\newblock \href {http://dx.doi.org/10.1061/41203(425)85}
  {\path{doi:10.1061/41203(425)85}}.

\bibitem{Tzatchkov_2008}
V.~G. Tzatchkov, V.~H. Alcocer-Yamanaka, V.~B. Ort{\'{\i}}z, Graph theory based
  algorithms for water distribution network sectorization projects, in: Water
  Distribution Systems Analysis Symposium 2006, American Society of Civil
  Engineers, 2008.
\newblock \href {http://dx.doi.org/10.1061/40941(247)172}
  {\path{doi:10.1061/40941(247)172}}.

\bibitem{Hajebi_2015}
S.~Hajebi, E.~Roshani, N.~Cardozo, S.~Barrett, A.~Clarke, S.~Clarke, Water
  distribution network sectorisation using graph theory and many-objective
  optimisation, Journal of Hydroinformatics{. }\href
  {http://dx.doi.org/10.2166/hydro.2015.144}
  {\path{doi:10.2166/hydro.2015.144}}.

\bibitem{liu_2016}
Y.-Y. Liu, A.-L. Barab{\'{a}}si, Control principles of complex systems, Reviews
  of Modern Physics 88~(3).
\newblock \href {http://dx.doi.org/10.1103/revmodphys.88.035006}
  {\path{doi:10.1103/revmodphys.88.035006}}.

\bibitem{Liu_2011}
Y.-Y. Liu, J.-J. Slotine, A.-L. Barab{\'{a}}si, Controllability of complex
  networks, Nature 473~(7346) (2011) 167--173.
\newblock \href {http://dx.doi.org/10.1038/nature10011}
  {\path{doi:10.1038/nature10011}}.

\bibitem{Ruths_2014}
J.~Ruths, D.~Ruths, Control profiles of complex networks, Science 343~(6177)
  (2014) 1373--1376.
\newblock \href {http://dx.doi.org/10.1126/science.1242063}
  {\path{doi:10.1126/science.1242063}}.

\bibitem{Summers_2014}
T.~H. Summers, J.~Lygeros, Optimal sensor and actuator placement in complex
  dynamical networks, {IFAC} Proceedings Volumes 47~(3) (2014) 3784--3789.
\newblock \href {http://dx.doi.org/10.3182/20140824-6-za-1003.00226}
  {\path{doi:10.3182/20140824-6-za-1003.00226}}.

\bibitem{yan_2012}
G.~Yan, J.~Ren, Y.-C. Lai, C.-H. Lai, B.~Li, Controlling complex networks: How
  much energy is needed?, Physical Review Letters 108~(21).
\newblock \href {http://dx.doi.org/10.1103/physrevlett.108.218703}
  {\path{doi:10.1103/physrevlett.108.218703}}.

\bibitem{yan_2015}
G.~Yan, G.~Tsekenis, B.~Barzel, J.-J. Slotine, Y.-Y. Liu, A.-L. Barab{\'{a}}si,
  Spectrum of controlling and observing complex~networks, Nature Physics 11~(9)
  (2015) 779--786.
\newblock \href {http://dx.doi.org/10.1038/nphys3422}
  {\path{doi:10.1038/nphys3422}}.

\bibitem{shirin_2017}
A.~Shirin, I.~S. Klickstein, F.~Sorrentino, Optimal control of complex
  networks: Balancing accuracy and energy of the control action, Chaos: An
  Interdisciplinary Journal of Nonlinear Science 27~(4) (2017) 041103.
\newblock \href {http://dx.doi.org/10.1063/1.4979647}
  {\path{doi:10.1063/1.4979647}}.

\bibitem{o_callaghan_1984}
J.~F. OCallaghan, D.~M. Mark, The extraction of drainage networks from digital
  elevation data, Computer Vision, Graphics, and Image Processing 27~(2) (1984)
  247.
\newblock \href {http://dx.doi.org/10.1016/s0734-189x(84)80047-x}
  {\path{doi:10.1016/s0734-189x(84)80047-x}}.

\bibitem{tarboton_1997}
D.~G. Tarboton, A new method for the determination of flow directions and
  upslope areas in grid digital elevation models, Water Resources Research
  33~(2) (1997) 309--319.
\newblock \href {http://dx.doi.org/10.1029/96wr03137}
  {\path{doi:10.1029/96wr03137}}.

\bibitem{shreve_1969}
R.~L. Shreve, Stream lengths and basin areas in topologically random channel
  networks, The Journal of Geology 77~(4) (1969) 397--414.
\newblock \href {http://dx.doi.org/10.1086/628366} {\path{doi:10.1086/628366}}.

\bibitem{rodriguez_2001}
I.~Rodriguez-Iturbe, A.~Rinaldo, Fractal river basins: chance and
  self-organization, Cambridge University Press, 2001.

\bibitem{tak_1990}
L.~D. Tak, R.~L. Bras, Incorporating hillslope effects into the geomorphologic
  instantaneous unit hydrograph, Water Resources Research 26~(10) (1990)
  2393--2400.

\bibitem{moore_1991}
I.~D. Moore, R.~Grayson, A.~Ladson, Digital terrain modelling: a review of
  hydrological, geomorphological, and biological applications, Hydrological
  processes 5~(1) (1991) 3--30.

\bibitem{this_repo_2018}
M.~Bartos, Controller placement code,
  \url{https://github.com/kLabUM/hydraulic-controller-placement} (2018).

\bibitem{pysheds_2018}
M.~Bartos, pysheds: simple and fast watershed delineation in python,
  \url{https://github.com/mdbartos/pysheds} (2018).

\bibitem{Lehner_2008}
B.~Lehner, K.~Verdin, A.~Jarvis, New global hydrography derived from spaceborne
  elevation data, Eos, Transactions American Geophysical Union 89~(10) (2008)
  93.
\newblock \href {http://dx.doi.org/10.1029/2008eo100001}
  {\path{doi:10.1029/2008eo100001}}.

\bibitem{nhd_2013}
{United States Geological Survey}, National hydrography geodatabase,
  \url{https://viewer.nationalmap.gov/viewer/nhd.html?p=nhd} (2013).

\bibitem{mays_2010}
L.~Mays, Water Resources Engineering, John Wiley \& Sons, 2010.

\bibitem{Moody_2002}
J.~A. Moody, B.~M. Troutman, Characterization of the spatial variability of
  channel morphology, Earth Surface Processes and Landforms 27~(12) (2002)
  1251--1266.
\newblock \href {http://dx.doi.org/10.1002/esp.403}
  {\path{doi:10.1002/esp.403}}.

\bibitem{swmm_2018}
{United States Environmental Protection Agency}, {ORD} stormwater management
  model, \url{https://github.com/USEPA/Stormwater-Management-Model} (2018).

\bibitem{Wong_2016}
B.~P. Wong, B.~Kerkez, Adaptive measurements of urban runoff quality, Water
  Resources Research 52~(11) (2016) 8986--9000.
\newblock \href {http://dx.doi.org/10.1002/2015WR018013}
  {\path{doi:10.1002/2015WR018013}}.

\bibitem{muschalla_2014}
D.~Muschalla, B.~Vallet, F.~Anctil, P.~Lessard, G.~Pelletier, P.~A.
  Vanrolleghem, Ecohydraulic-driven real-time control of stormwater basins,
  Journal of Hydrology 511 (2014) 82--91.
\newblock \href {http://dx.doi.org/10.1016/j.jhydrol.2014.01.002}
  {\path{doi:10.1016/j.jhydrol.2014.01.002}}.

\bibitem{poresky_2015}
A.~Poresky, R.~Boyle, O.~Cadwalader, Piloting real time control retrofits of
  stormwater facilities: Two oregon case studies and beyond, Proceedings of the
  Pacific Northwest Clean Water Association, Boise, ID, USA (2015) 26--27.

\end{thebibliography}

\clearpage

\part*{Supplementary information}

\setcounter{figure}{0}
\setcounter{section}{0}
\renewcommand{\thefigure}{S\arabic{figure}}
\renewcommand{\thetable}{S\arabic{table}}
\renewcommand{\thesection}{S\arabic{section}}

\section{Implementations of algorithms used in the study}

\subsection{Width function}

For the software implementation used in this study, the width function is
computed by determining the travel times from each vertex to the outlet, and
then generating a binned histogram of these travel times. The travel times from
each vertex to the outlet are computed by performing a depth-first search on the
graph representation of the watershed starting with the outlet, and then
recording the distances from each vertex to the outlet. The travel times are
then binned to produce the width function. The travel time computation is
implemented as \textbf{grid.flow\_distance} in the \textit{pysheds} software
package, available at \url{github.com/mdbartos/pysheds}.

Note that if matrix multiplication is used to compute the width function, the
inter-vertex travel times cannot be used as the weights of the adjacency matrix.
Rather, differential travel times may be accounted for by modifying the topology
of the graph. For instance, consider a graph consisting of fast nodes and slow
nodes, where fast nodes transfer flow 10 times as quickly as slow nodes. In this
scheme, slow nodes can be modeled using 10 ``dummy'' vertices placed in series.
It should be noted however, that this implementation is inefficient both in
terms of speed of computation and memory usage.

\clearpage

\subsection{Flow accumulation}

The vectorized flow accumulation algorithm (developed previously by the authors,
but unpublished) is reproduced here for the reader's convenience:

\begin{small}
\begin{enumerate}
\item Create an m x n array, \textbf{edges} to represent the edges of the
  directed graph. For each entry in the array,
  the index corresponds to the index of the start node, and the value
  corresponds to the index of the end node.
\item Create an m x n array, \textbf{in\_degree}, to hold the in-degree
  of each grid cell (i.e. the number of cells currently pointing to that cell).
  This can be accomplished by counting the number of occurrences of each unique
  value in \textbf{edges}.
\item Create an m x n array of ones \textbf{flow\_accumulation} to hold the
  computed number of upstream cells for each cell.
\item Define a 1 x nm array \textbf{startnodes} with entries equal to the
  indices of \textbf{edges}.
\item Define a 1 x nm array \textbf{endnodes} with entries equal to the values
  of \textbf{edges}.
\item Create a 1 x nm boolean index \textbf{no\_predecessors} which is 0 where
  \textbf{in\_degree} is greater than 0, and 1 where \textbf{in\_degree} is
  equal to zero.
\item Select the subset of start nodes and end nodes that have no predecessors:
  \textbf{startnodes = startnodes[no\_predecessors]} and \textbf{endnodes =
    endnodes[no\_predecessors]}. This selects the ``outermost layer'' of nodes.
\item While \textbf{endnodes} is not empty:
  \begin{itemize}
  \item Add the flow accumulation at the start nodes to the flow accumulation of
    the end nodes: \textbf{flow\_accumulation[endnodes] += flow\_accumulation[startnodes]}
  \item Decrement the in-degree of the endnodes by the number of start nodes
    that are linked to it in the current step. With \textbf{endnodes} containing
    repeated entries this operation can be represented as: \textbf{in\_degree[endnodes] -=
      1}.
  \item Set the new value of \textbf{startnodes} as the unique elements in
    \textbf{endnodes} with a current in-degree of zero: \textbf{startnodes =
      unique(endnodes)[(in\_degree == 0)]}
  \item Set \textbf{endnodes} as the end nodes corresponding to the new start nodes: \textbf{endnodes = edges[startnodes]}.
  \end{itemize}
\end{enumerate}
\end{small}

This algorithm is implemented as \textbf{grid.accumulation} in the
\textit{pysheds} software package, available at
\url{github.com/mdbartos/pysheds}.

\clearpage

\section{Vertex weights as a function of $\phi$}

\begin{figure*}[htb!]
\centering
\begin{subfigure}[b]{\textwidth}
   \includegraphics[width=\textwidth]{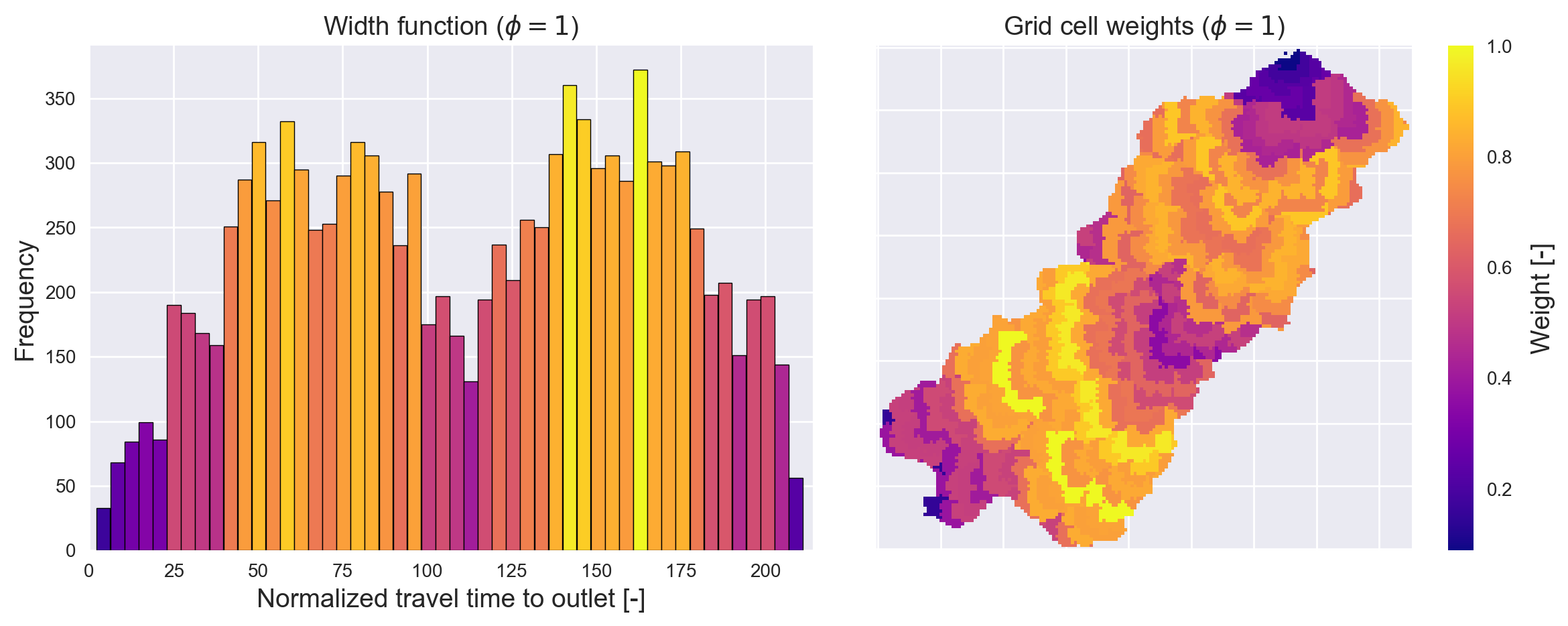}
\end{subfigure}

\begin{subfigure}[b]{\textwidth}
   \includegraphics[width=\textwidth]{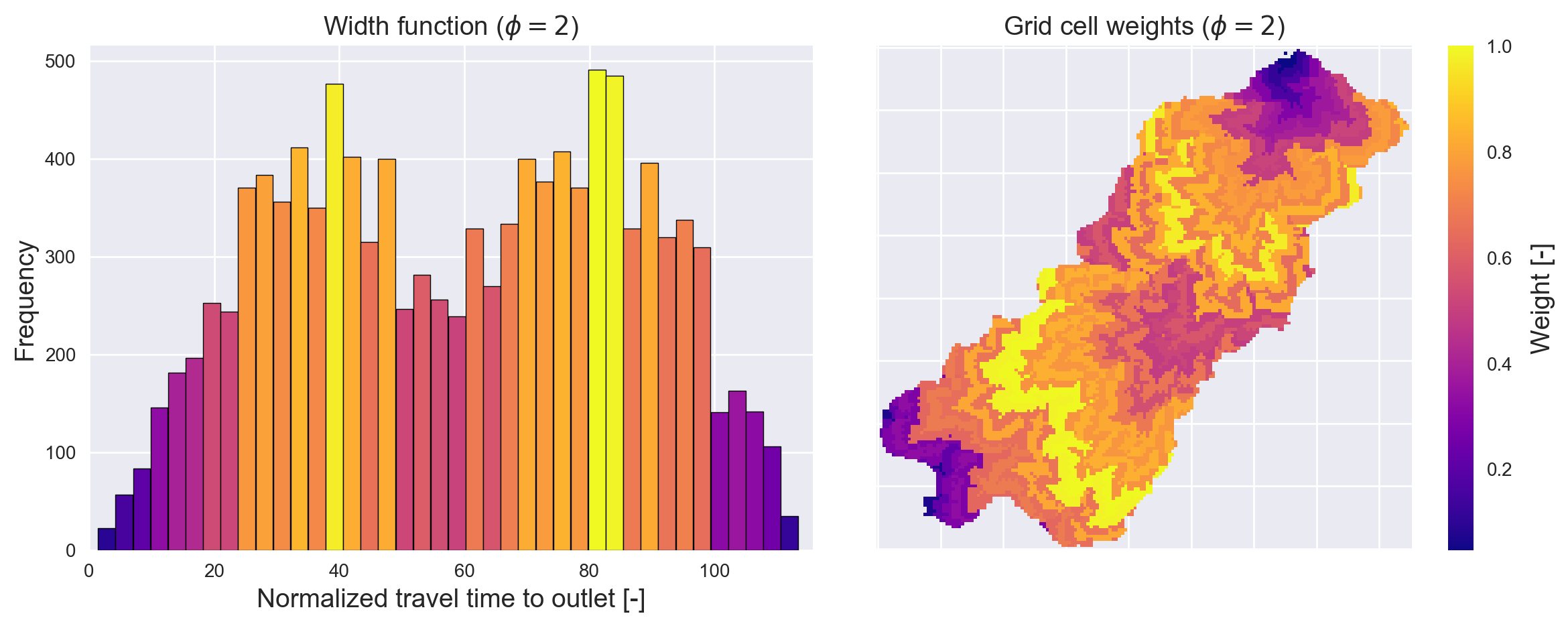}
\end{subfigure}

\begin{subfigure}[b]{\textwidth}
   \includegraphics[width=\textwidth]{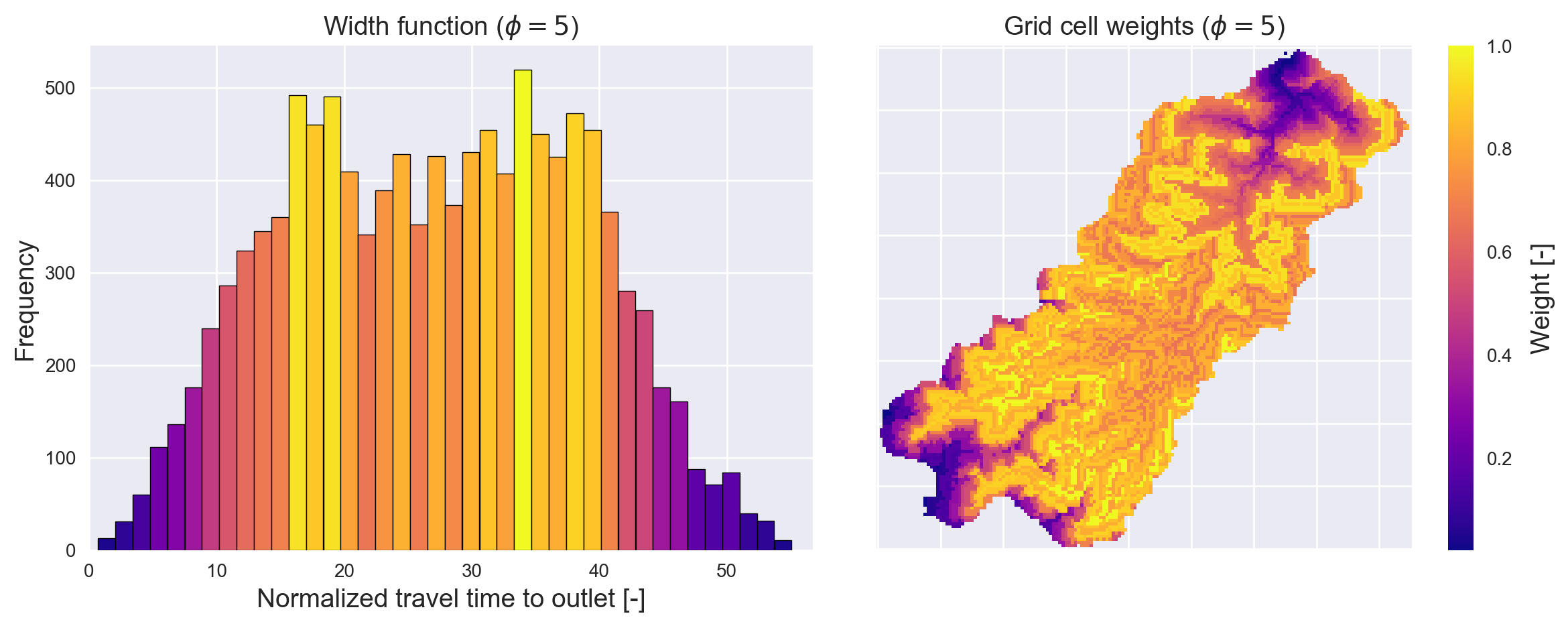}
\end{subfigure}

\caption[Vertex weights as a function of $\phi$, $\phi \in \{ 1, 2, 5\}$]{Width
  function (left) and vertex weights (right) as a function of $\phi$, with $\phi \in \{ 1, 2, 5\}$}
\label{fig:weights_1}
\end{figure*}

\clearpage

\begin{figure*}[htb!]
\centering
\begin{subfigure}[b]{\textwidth}
   \includegraphics[width=\textwidth]{img/weights_phi10.png}
\end{subfigure}

\begin{subfigure}[b]{\textwidth}
   \includegraphics[width=\textwidth]{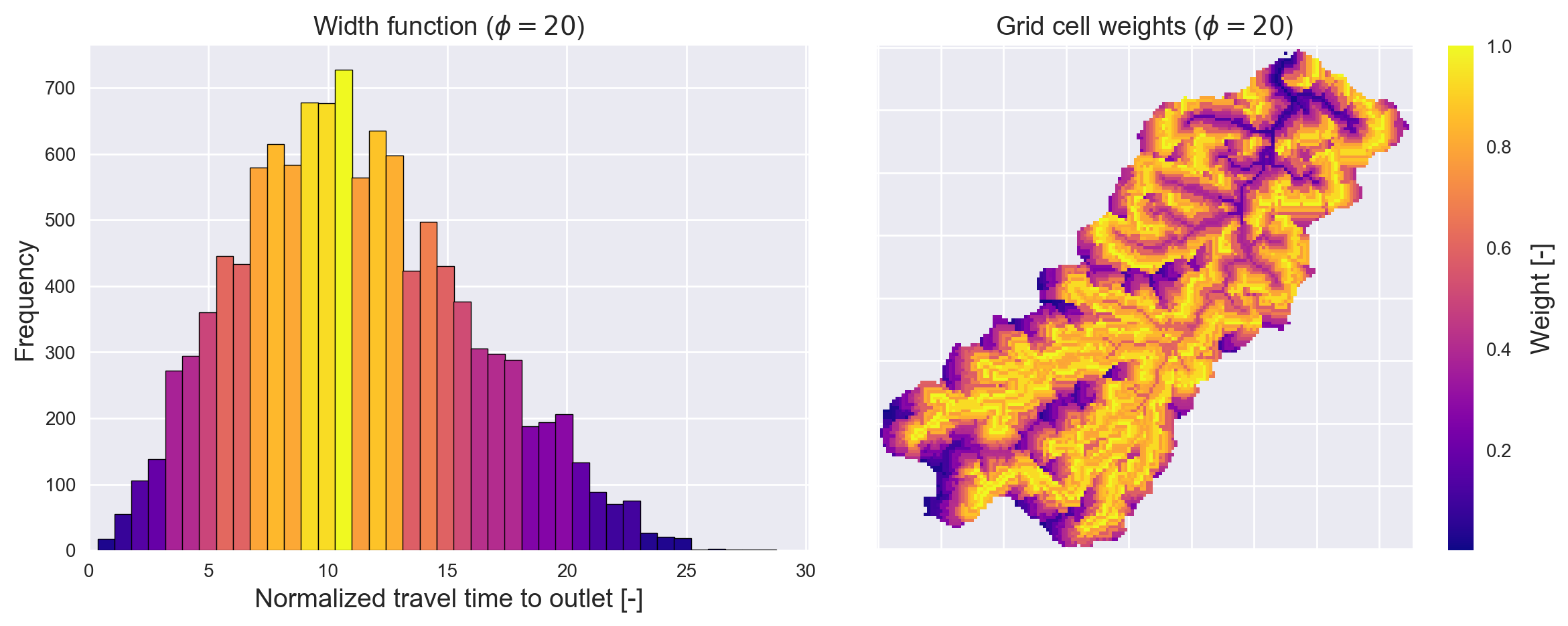}
\end{subfigure}

\begin{subfigure}[b]{\textwidth}
   \includegraphics[width=\textwidth]{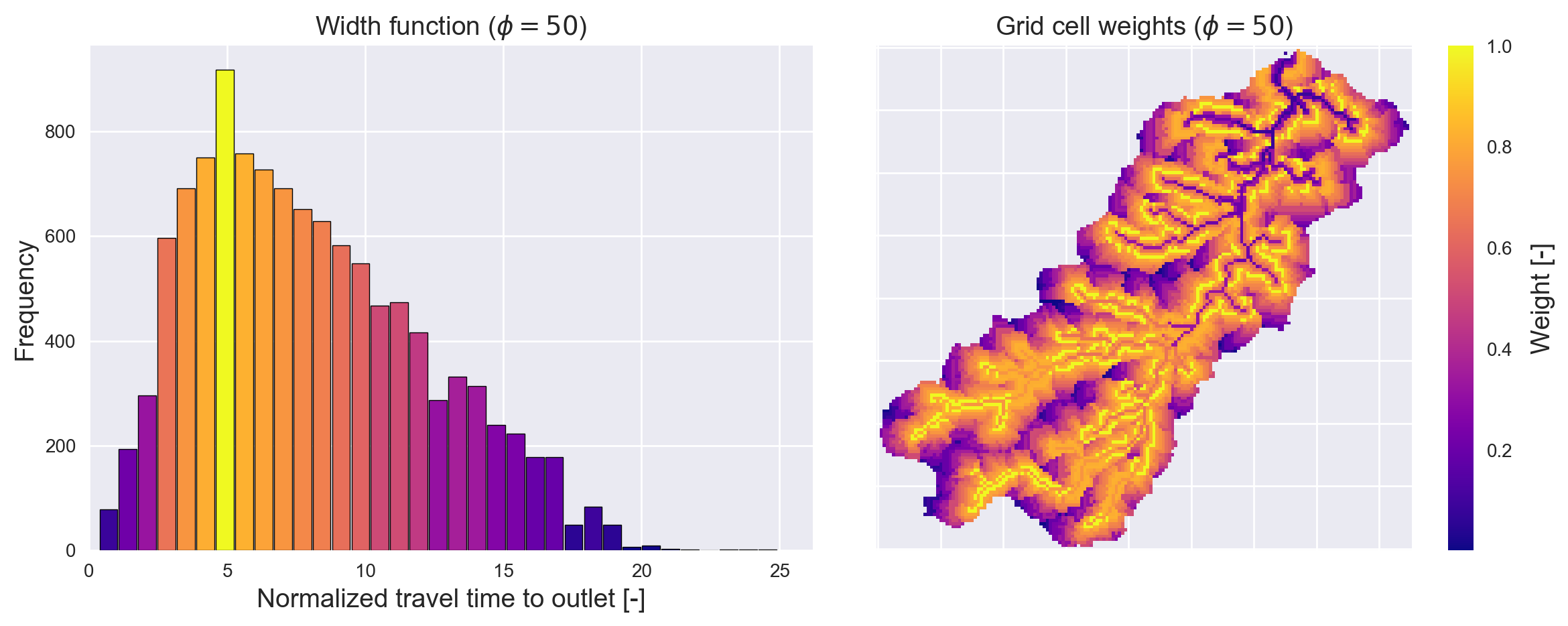}
\end{subfigure}

\caption[Vertex weights as a function of $\phi$, $\phi \in \{ 10, 20,
50\}$]{Width function (left) and vertex weights (right) as a function of $\phi$,
  with $\phi \in \{ 10, 20, 50\}$}
\label{fig:weights_2}
\end{figure*}

\clearpage

\section{Optimal controller placements for various numbers of controllers ($k$)}

\begin{figure*}[htb!]
\centering
\begin{subfigure}[b]{\textwidth}
  \includegraphics[width=\textwidth]{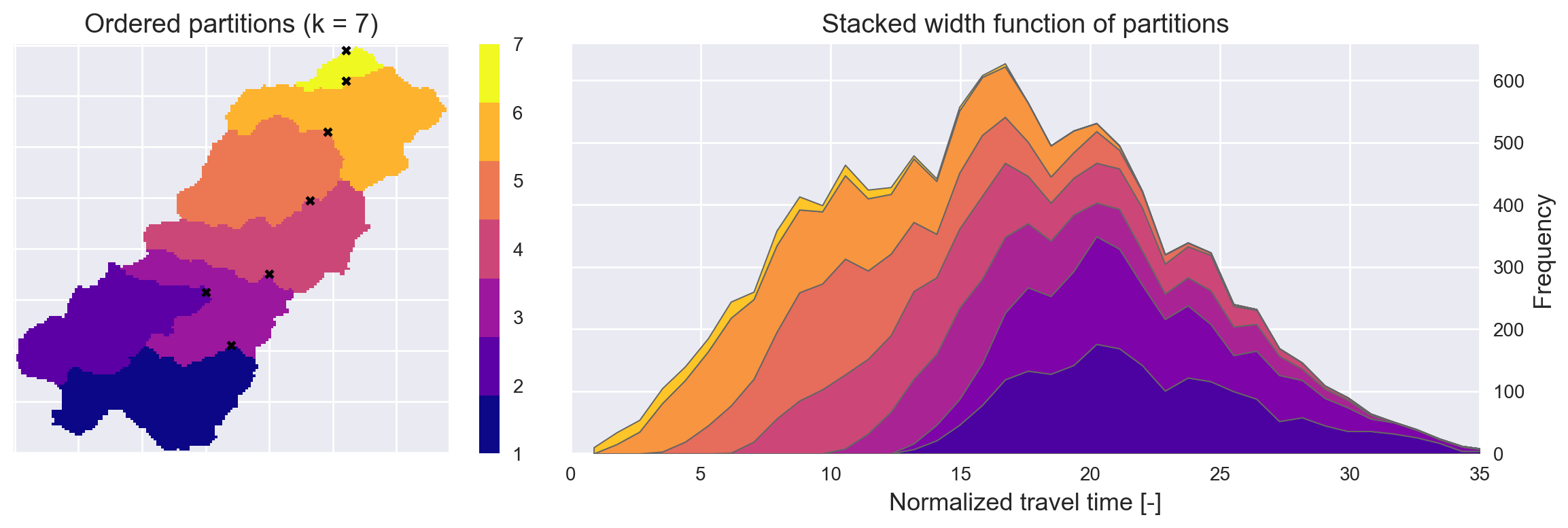}
\end{subfigure}

\begin{subfigure}[b]{\textwidth}
   \includegraphics[width=\textwidth]{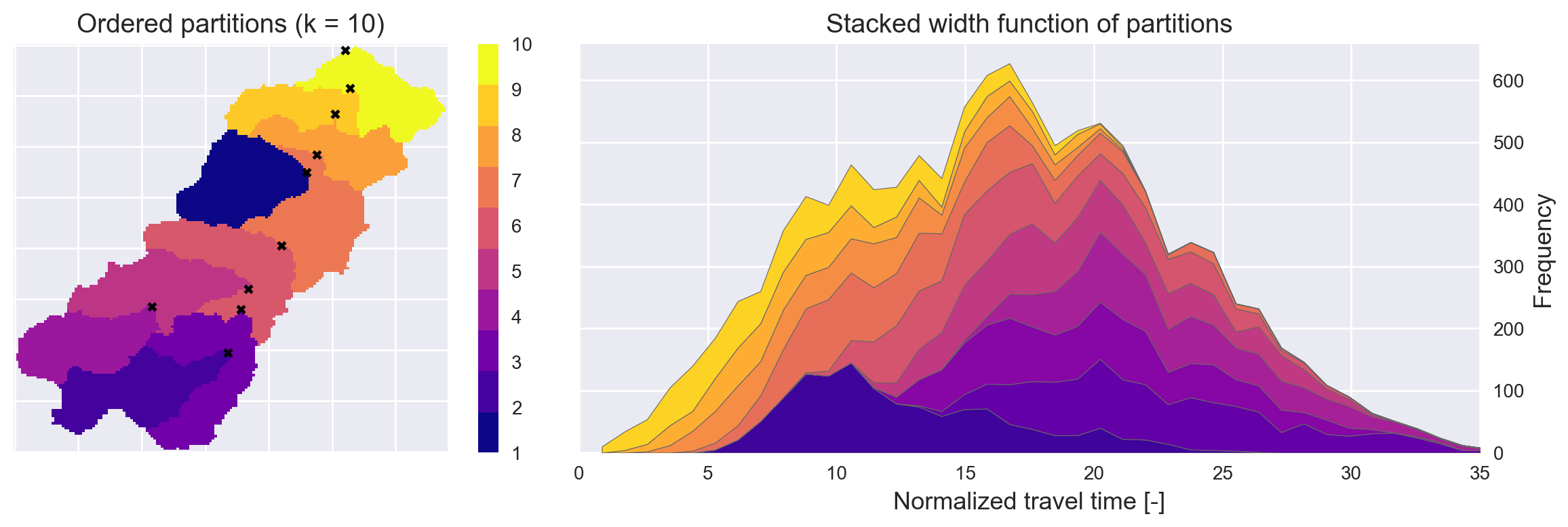}
\end{subfigure}

\begin{subfigure}[b]{\textwidth}
   \includegraphics[width=\textwidth]{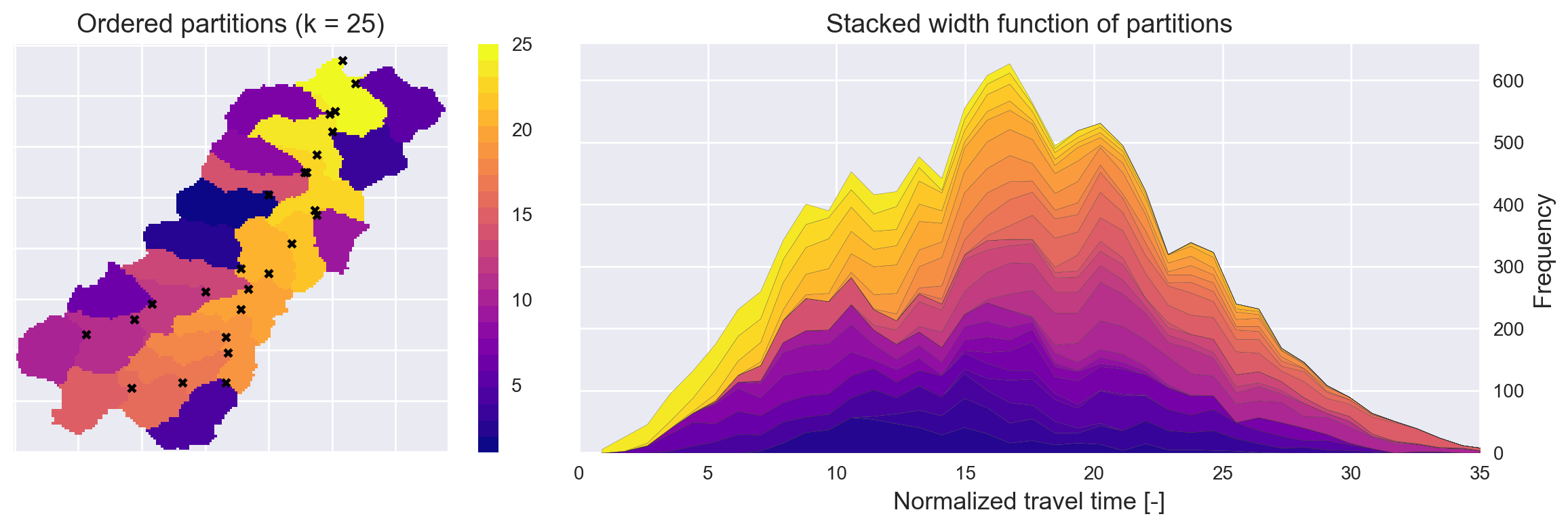}
\end{subfigure}

\caption[Partitions and stacked width functions for varying numbers of
controllers]{Optimized controller placements (left) and stacked width functions
  for varying number of controllers (right), $k \in \{ 7, 10, 25\}$, $\phi=10$.}
\label{fig:placements}
\end{figure*}

\clearpage

\section{Performance under rainfall events of different sizes}

\begin{figure*}[htb!]
\centering
\begin{subfigure}[b]{\textwidth}
   \includegraphics[width=\textwidth]{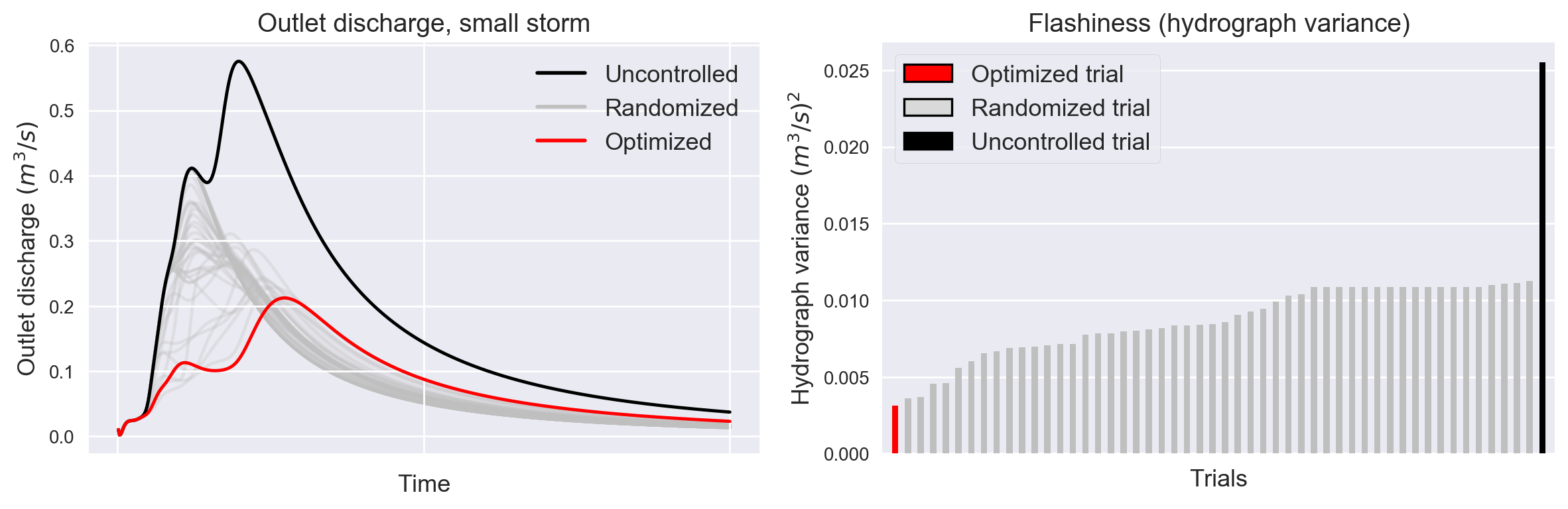}
\end{subfigure}

\begin{subfigure}[b]{\textwidth}
   \includegraphics[width=\textwidth]{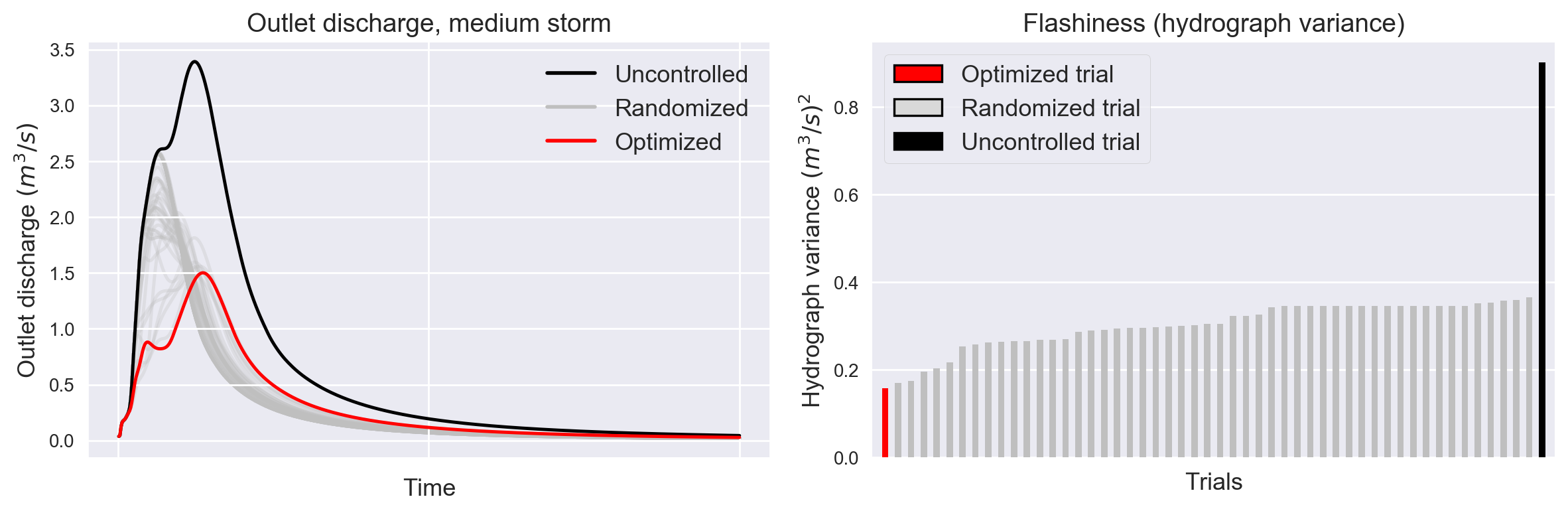}
\end{subfigure}

\begin{subfigure}[b]{\textwidth}
   \includegraphics[width=\textwidth]{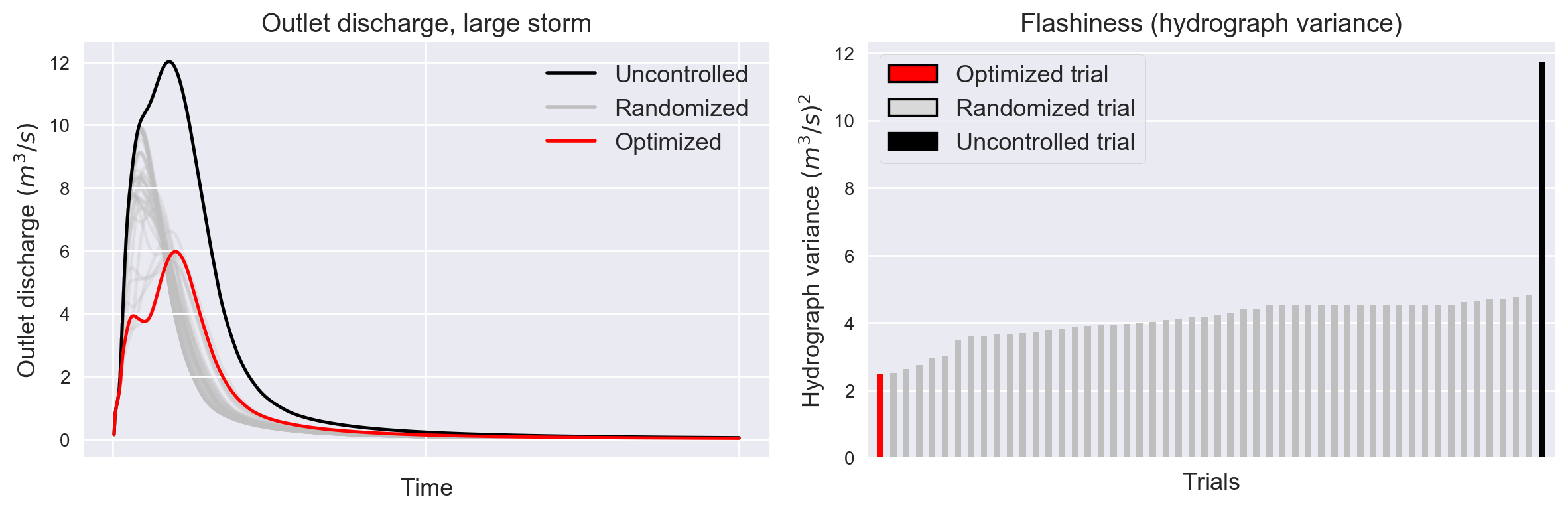}
\end{subfigure}
\caption[Hydraulic modeling results]{Left: simulated hydrographs for the
  uncontrolled scenario (black), naive controller placement (gray), and the
  optimized controller placement (red) under small, medium and large storm
  events (top to bottom). Right: flashiness (as measured by the variance of the
  hydrograph) for each simulation.}
\label{fig:variance}
\end{figure*}

\begin{figure*}[htb!]
\centering
\begin{subfigure}[b]{\textwidth}
   \includegraphics[width=\textwidth]{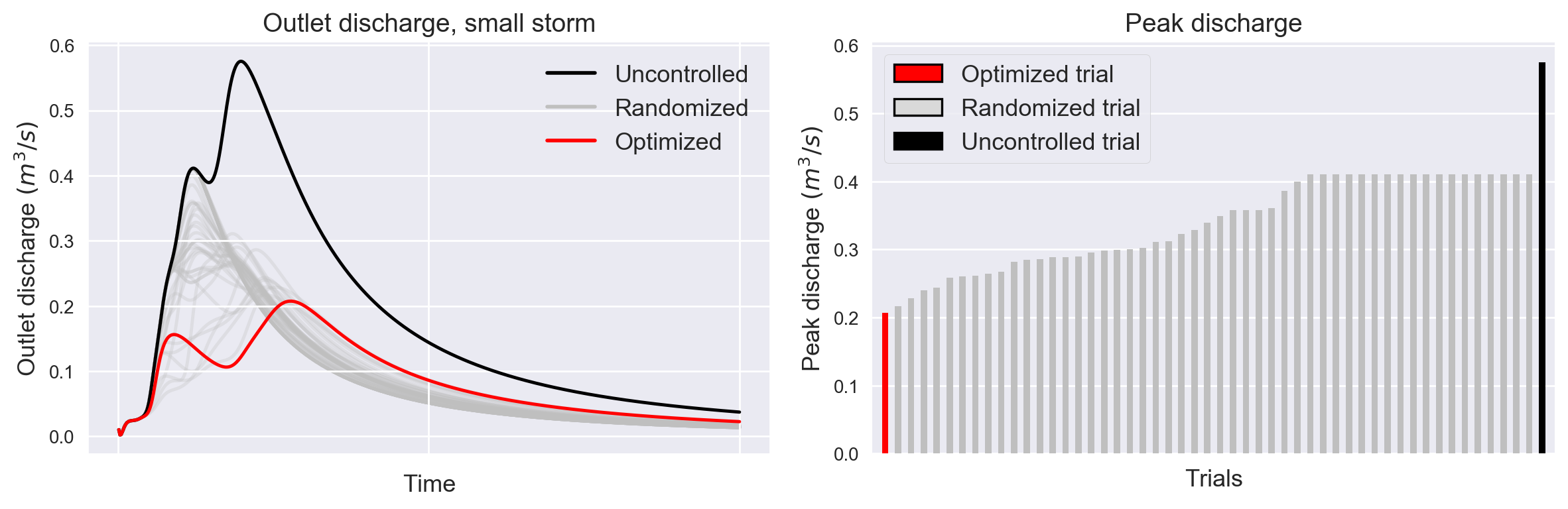}
\end{subfigure}

\begin{subfigure}[b]{\textwidth}
   \includegraphics[width=\textwidth]{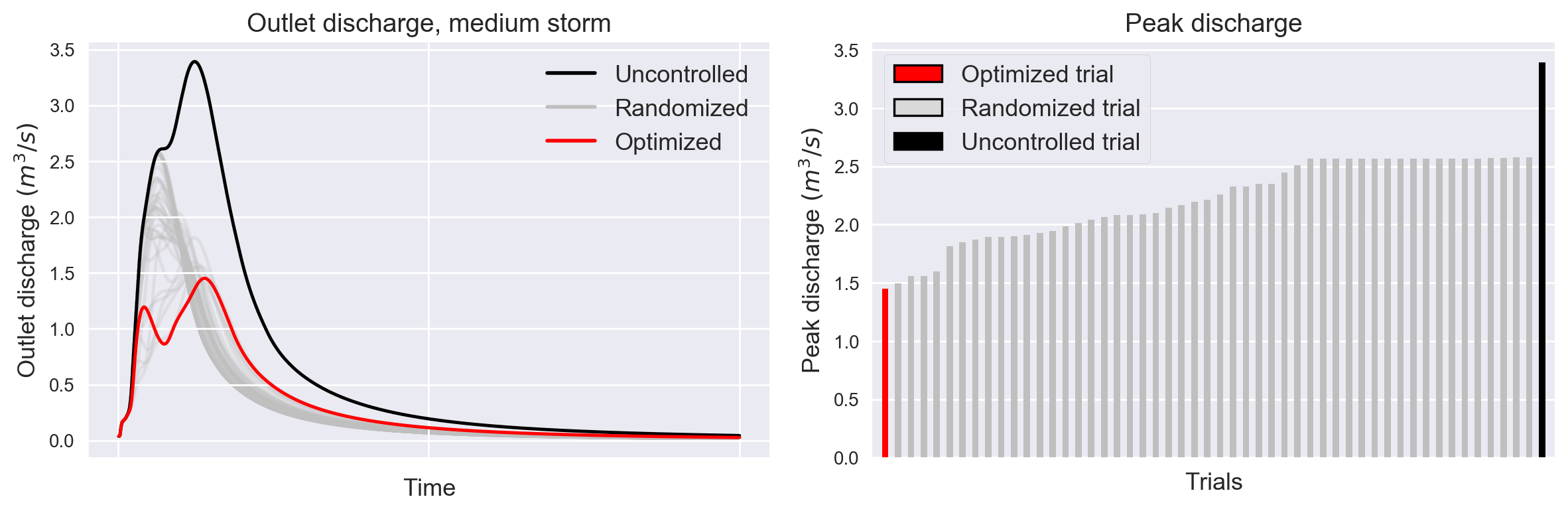}
\end{subfigure}

\begin{subfigure}[b]{\textwidth}
   \includegraphics[width=\textwidth]{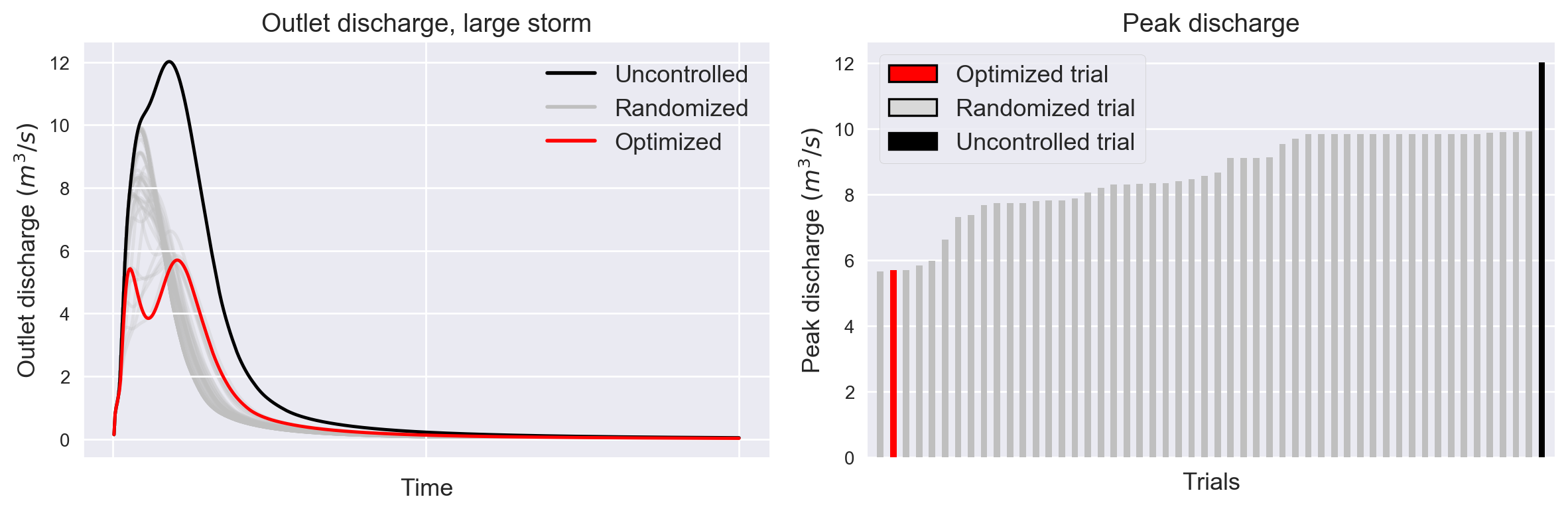}
\end{subfigure}
\caption[Hydraulic modeling results]{Left: simulated hydrographs for the
  uncontrolled scenario (black), naive controller placement (gray), and the
  optimized controller placement (red) under small, medium and large storm
  events (top to bottom). Right: peak discharge for each simulation.}
\label{fig:peak}
\end{figure*}

\begin{figure*}[htb!]
\centering
\begin{subfigure}[b]{\textwidth}
   \includegraphics[width=\textwidth]{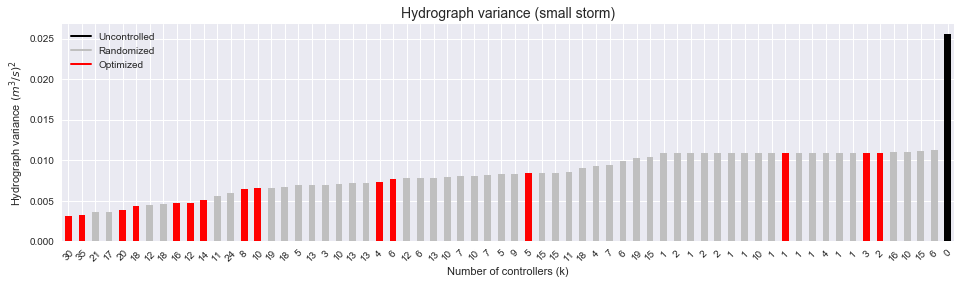}
\end{subfigure}

\begin{subfigure}[b]{\textwidth}
   \includegraphics[width=\textwidth]{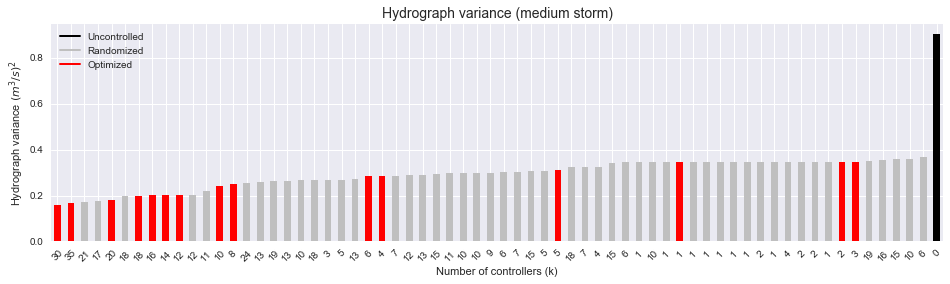}
\end{subfigure}

\begin{subfigure}[b]{\textwidth}
   \includegraphics[width=\textwidth]{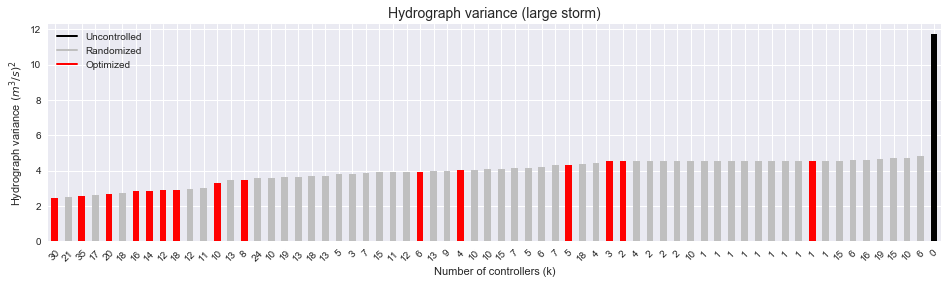}
\end{subfigure}
\caption[Hydraulic modeling results]{Hydrograph variance for small, medium and
  large storms under all model runs.}
\label{fig:var_all}
\end{figure*}

\begin{figure*}[htb!]
\centering
\begin{subfigure}[b]{\textwidth}
   \includegraphics[width=\textwidth]{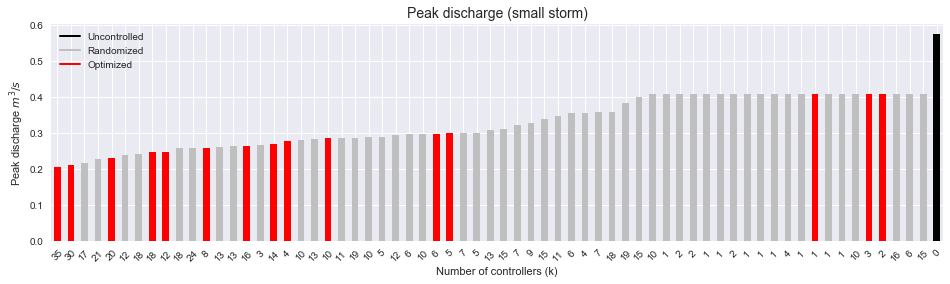}
\end{subfigure}

\begin{subfigure}[b]{\textwidth}
   \includegraphics[width=\textwidth]{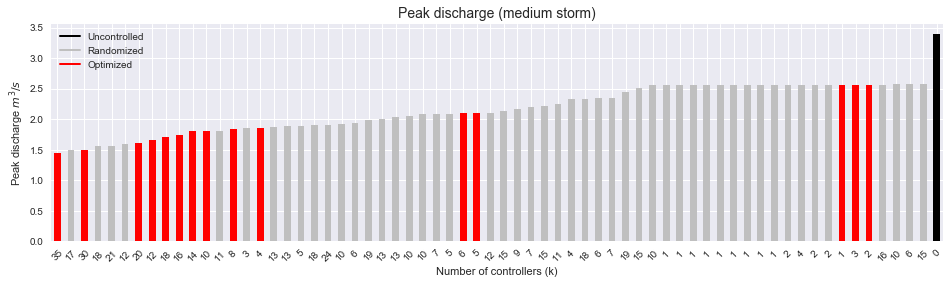}
\end{subfigure}

\begin{subfigure}[b]{\textwidth}
   \includegraphics[width=\textwidth]{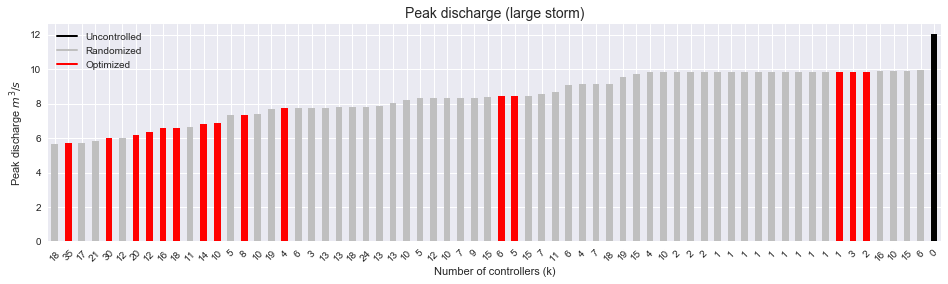}
\end{subfigure}
\caption[Hydraulic modeling results]{Peak discharge for small, medium and
  large storms under all model runs.}
\label{fig:peak_all}
\end{figure*}

When tested against storm events of different sizes, the controller placement
algorithm generally outperforms randomized control trials. However, the relative
performance between simulations varies with rainfall intensity, which suggests
that a uniquely optimal controller placement may not exist for rainfall events
of all sizes. As shown in Figures \ref{fig:variance} and \ref{fig:peak}, the
optimized controller placement still produces flatter flows overall compared to
randomized trials for the small and large storm events. However, for the large
storm event, one of the randomized simulations produces a slightly smaller peak
discharge than the best-performing optimized controller placement. Moreover, the
within-group performance of controller placement strategies varies with storm
event size, as seen in Figures \ref{fig:var_all} and \ref{fig:peak_all}. For
instance, the controller placement that produces the smallest peak discharge
under the large storm event produces the 4th smallest peak discharge under the
medium storm event, and the 7th smallest peak discharge under the small storm
event (Figure \ref{fig:peak_all}). These results suggest that the optimal
controller placement for large storms may not be the same as the optimal
controller placement for small storms. This situation may result from the fact
that larger flood waves travel faster, meaning that inter-vertex travel times
will change depending on the scale of the hydrologic response. Consequently,
assumed inter-vertex travel times (controlled in this experiment by the
parameter $\phi$) may need to be tuned depending on storm event size to account
for the nonlinearities inherent in flood wave travel times. Despite these
parameter selection issues, the experiments show that controller placement
algorithm still produces flatter flows than random controls for storm events of
various sizes.

\clearpage

\section{Hydrograph variance for storm events of different sizes}

\clearpage

\section{Peak discharge for storm events of different sizes}

\clearpage

\section{Hydrograph variance for all simulations}

\clearpage

\section{Peak discharge for all simulations}

\clearpage

\section{Performance metrics by number of controllers}

\begin{figure*}[htb!]
\centering
\begin{subfigure}[b]{\textwidth}
   \includegraphics[width=\textwidth]{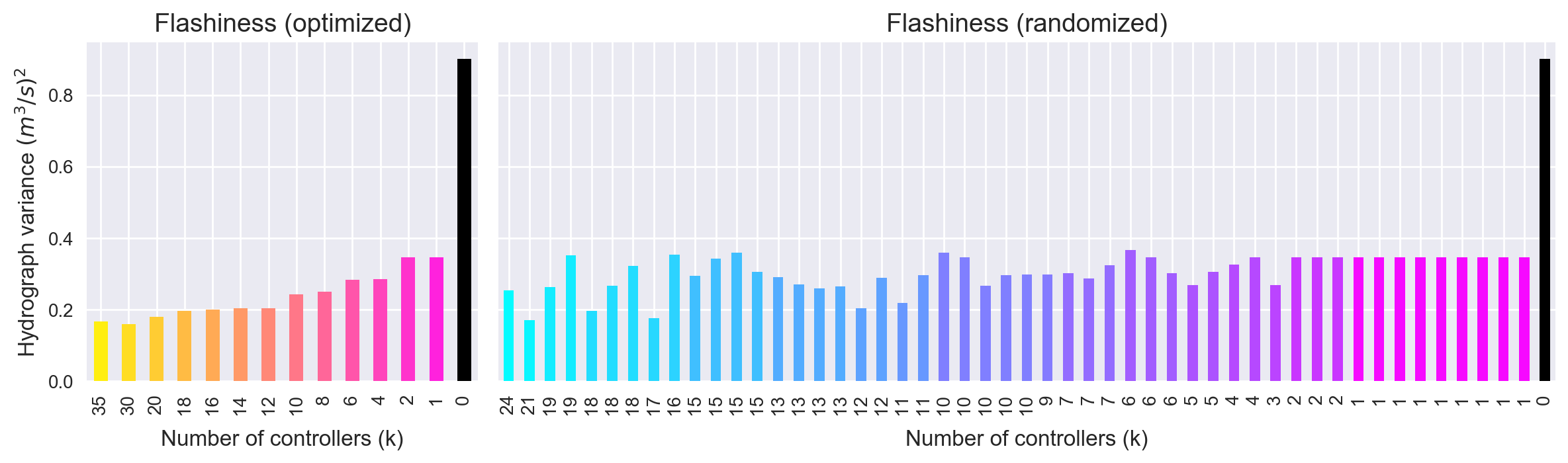}
\end{subfigure}

\begin{subfigure}[b]{\textwidth}
   \includegraphics[width=\textwidth]{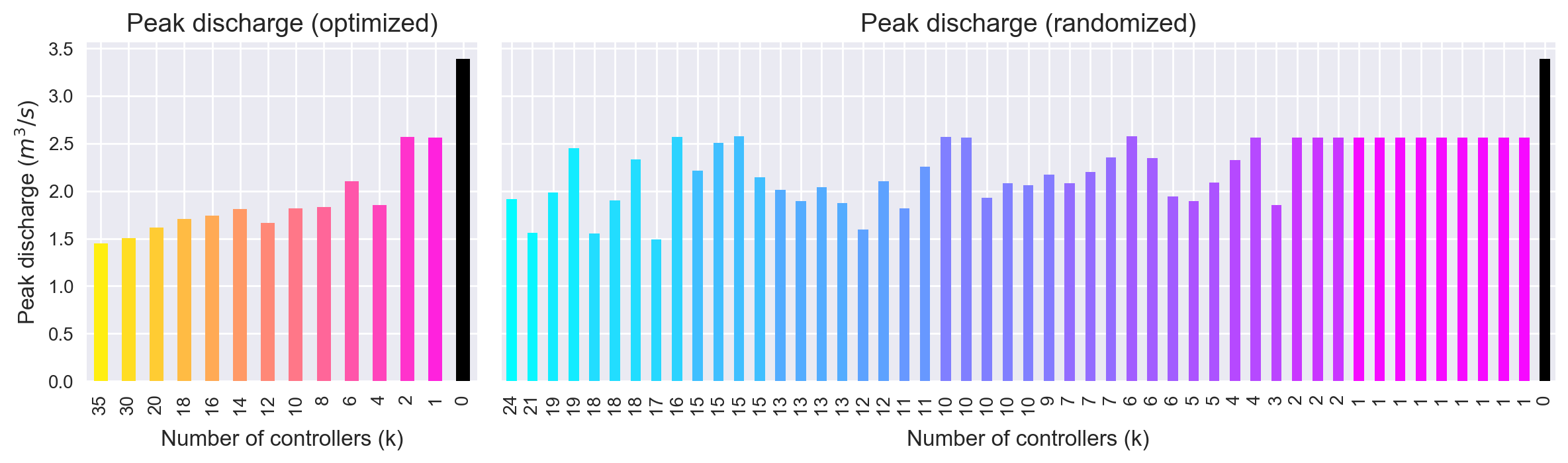}
\end{subfigure}

\caption[Hydraulic modeling results]{Performance metrics for both optimized
  (left) and randomized (right) controller placements by number of controllers
  used. The top panel measures performance in terms of flashiness (hydrograph
  variance), while the bottom panel measures performance by peak discharge. The
  optimized controller placements show consistently better performance metrics
  as the number of controllers is increased, while the randomized simulations do
  not.}
\label{fig:num_controllers_comparison}
\end{figure*}

\end{document}